\documentclass[journal=jacsat,manuscript=article]{achemso}
\usepackage{graphicx}
\usepackage{booktabs}

\usepackage{dcolumn}
\usepackage{enumitem} 

\usepackage{bm}
\usepackage{mathrsfs}
\usepackage{amsmath}
\usepackage{mathtools}
\usepackage{physics}
\usepackage{braket}
\usepackage{siunitx}

\usepackage{doi}

\DeclareSIUnit{\atomicunit}{a.u.}
\DeclareSIUnit{\electronvolt}{eV}

\usepackage{caption}
\usepackage[font={footnotesize}]{caption}

\usepackage[]{hyperref}
\usepackage{cleveref}

\usepackage[]{xcolor}
\usepackage{soul}

\usepackage{pdfpages}

\author{Yassir El Moutaoukal}
\affiliation{Department of Chemistry, Norwegian University of Science and Technology, 7491 Trondheim, Norway}
\alsoaffiliation{These authors contributed equally to this work}
\author{Rosario R. Riso}
\affiliation{Department of Chemistry, Norwegian University of Science and Technology, 7491 Trondheim, Norway}
\alsoaffiliation{These authors contributed equally to this work}
\author{Matteo Castagnola}
\affiliation{Department of Chemistry, Norwegian University of Science and Technology, 7491 Trondheim, Norway}
\author{Enrico Ronca}
\affiliation{Department of Chemistry, Biology and Biotechnology, University of Perugia, Via Elce di Sotto, 8, 06123, Perugia, Italy}
\author{Henrik Koch}
\affiliation{Department of Chemistry, Norwegian University of Science and Technology, 7491 Trondheim, Norway}
\email{henrik.koch@ntnu.no}
\title[]
  {\Large{Strong coupling M{\o}ller-Plesset perturbation theory}}
  
\abbreviations{QED, Perturbation theory, M{\o}ller-Plesset, Molecular orbitals, Polaritons, electron-photon correlation, Strong coupling}
\keywords{QED, Perturbation theory, M{\o}ller-Plesset, Molecular orbitals, Polaritons, electron-photon correlation, Strong coupling}

\begin{document}

\begin{abstract}
Perturbative approaches are methods to efficiently tackle many-body problems, offering both intuitive insights and analysis of correlation effects. However, their application to systems where light and matter are strongly coupled is non-trivial. Specifically, the definition of suitable orbitals for the zeroth-order Hamiltonian represents a significant theoretical challenge. While reviewing previously investigated orbital choices, this work presents an alternative polaritonic orbital basis suitable for the strong coupling regime. We develop a quantum electrodynamical (QED) M{\o}ller-Plesset perturbation theory using orbitals obtained from the strong coupling QED Hartree-Fock. We assess the strengths and limitations of the different approaches and emphasize the essential role of using a consistent molecular orbital framework to achieve an accurate description of cavity-induced electron-photon correlation effects.
\end{abstract}

\section{1. Introduction}
Strong coupling between electromagnetic vacuum fluctuations and matter allows for non-invasive engineering of molecular properties.~\cite{garcia2021manipulating,flick2018strong,herrera2020molecular,sandik2024cavity,fukushima2022inherent,schachenmayer2015cavity}
To achieve such a regime, experimentalists couple molecules with optical devices able to confine the electromagnetic fields in small quantization volumes.~\cite{mondal2022strong,fitzgerald2016quantum,hubener2021engineering} 
When molecular excitations interact with the quantized fields, hybrid states called molecular polaritons are formed.~\cite{scholes2020polaritons,mauro2024classical,sasaki2025optical}
Such states display unique features that can be adjusted by tuning the properties of the quantized field.~\cite{ribeiro2018polariton} Potential applications of polaritonic chemistry range from the modification of molecular absorption and emission spectra to the potential catalysis of chemical reactions.~\cite{fregoni2020strong,lather2019cavity,ahn2023modification,mony2021photoisomerization,mandal2023theory,fischer2022cavity,krupp2024quantum} 
While experimental efforts keep advancing into the ultrastrong coupling regime, a theoretical comprehension of the experimental results is still necessary.~\cite{chikkaraddy2016single,santhosh2016vacuum,bitton2022plasmonic,baumberg2022picocavities,damari2019strong,castagnola2024collective}
Modeling the light-matter interplay requires the use of QED theory in  order to capture the correlation effects between electrons and photons.~\cite{ashida2021cavity,haugland2020coupled,taylor2020resolution,mctague2022non,du2022catalysis,huang2025analytical}
Several quantum chemistry \textit{ab initio} methodologies have been extended to QED environments, such as the quantum electrodynamical density functional theory (QEDFT)\cite{ruggenthaler2014quantum,schafer2022shining,schafer2018ab,flick2018ab} and the quantum electrodynamical coupled cluster (QED-CC).~\cite{riso2022characteristic,pavosevic2022cavity,deprince2021cavity,liebenthal2022equation,mordovina2020polaritonic}
Despite its computational affordability, the accuracy of QEDFT relies on an electron-photon correlation functional that is challenging to model,~\cite{flick2018ab} while the more accurate QED-CC exhibits a steep computational scaling with system size.~\cite{helgaker2013molecular}
Perturbative methodologies are reliable alternatives to estimate correlation at a cheaper computational cost while providing, at the same time, an intuitive understanding of the complex interplay between the components of the many-body system.
Inside an optical cavity, perturbative approaches can either be obtained by excluding the field-dependent terms from the unperturbed Hamiltonian, in line with what is reported by Haugland \textit{et al.},~\cite{haugland2023perturbation} or by retaining the mean-field effects of the cavity in the zeroth-order Hamiltonian.
Bauer \textit{et al.}~\cite{bauer2023perturbation} reported an implementation of the first QED versions of the second order M{\o}ller-Plesset methodology (MP2) and the algebraic diagrammatic construction for the polarization propagator (ADC(2)). 
Specifically, the method are built starting from two possible reference states: the standard non-polaritonic  Hartree-Fock state (QED(np-HF)-MP2 and QED(np-HF)-ADC(2)), and the QED Hartree-Fock (QED-HF) wave function (QED-MP2 and QED-ADC(2)).
These approaches seem to accurately describe light-matter states while incorporating a significant part of many-body correlation.
These are surprising findings as Haugland \textit{et al.}~\cite{haugland2020coupled} demonstrated that the QED-HF molecular orbitals display unphysical properties, such as their lack of origin invariance for charged molecular systems due to an incorrect construction of the QED-Fock operator posing issues when developing post-HF perturbation theories.

The polaritonic molecular orbital problem for QED environments was addressed by Riso \textit{et al.}~\cite{riso2022molecular} introducing a novel \textit{ab initio} approach named strong coupling quantum electrodynamics Hartree-Fock (SC-QED-HF) theory.
This model not only provides fully consistent molecular orbitals by dressing the electrons with the cavity photons, but is also able to capture to some extent electron-photon correlation already at the mean-fied level.
Recent improvements in the convergence of SC-QED-HF by means of second order algorithms\cite{el2024toward} prompted us to develop of a M{\o}ller-Plesset perturbation theory starting from this alternative reference wave function.
We denote this method as SC-QED M{\o}ller-Plesset perturbation theory.
Another perturbative approach based on a wave function parametrization similar to the SC-QED-HF was recently published, namely the Lang-Firsov M{\o}ller-Plesset scheme (LF-MP2).~\cite{cui2024variational} 
The main difference between the two methods is that the diagonal Lang-Firsov transformation in LF-MP2 is not performed in a basis that diagonalizes the dipole operator.
Our findings demonstrate that the developed SC-QED-MP2 accurately reproduces the field-induced electron-photon correlation effects by capturing them already at the mean-field level.
This is especially the case as the light-matter coupling increases due to the exactness of the reference wave function in the infinite coupling limit.
Moreover, because of the non size-intensivity of the zeroth-order Hamiltonian in QED-MP2 theory, unphysical behavior in the long-range regime between two molecular systems is observed. The same issue emerges for the LF-MP2 method too suggesting that the choice of basis in QED \textit{ab initio} approaches is a delicate matter.

This paper is organized as follows. In Section 2, we describe the different choices for the zeroth-order Hamiltonian, deriving the energy expressions for QED-MP2, QED(np-HF)-MP2, and SC-QED-MP2 theory. In this Section we also show the differences between the SC-QED-MP2 and LF-MP2 approaches. In Section 3, we compare the performance of methodologies, focusing on coupling and frequency dispersions, intermolecular interactions, and cavity polarization orientational effects. Finally, our concluding remarks are presented in Section 4.

\section{2. Theory}
The interaction between light and matter inside an optical cavity can be modeled using the single-mode Pauli-Fiertz Hamiltonian in the dipole approximation and length gauge\cite{cohen2024photons,mandal2020polarized,de2014light,di2019resolution,frisk2019ultrastrong}
\begin{equation}\label{eq:Lenght_gauge}
    \begin{split}
        H &= H_{e} + \omega  b^\dagger b + \frac{\lambda^2}{2}  ( {\mathbf{d}} \cdot \pmb{\epsilon})^2 \\
        & - \lambda\sqrt{\frac{\omega}{2}} (    {\mathbf{d}}\cdot \pmb{\epsilon}) ( b^{\dagger}+b ) ,
    \end{split}
\end{equation}
where $b$ and $b^{\dagger}$ annihilate and create photons of frequency $\omega$ and with polarization $\pmb{\epsilon}$. The $\lambda$ parameter represents the light-matter coupling strength for a field confinement volume $V$ and  relative permittivity for the medium within the cavity $\epsilon_r$
\begin{equation}
    \lambda= \sqrt{\frac{2\pi}{\epsilon_r V}} ,
\end{equation}
while $\mathbf{d}$ is the molecular dipole operator\cite{haugland2020coupled}
\begin{equation}
\mathbf{d}= \sum_{pq}\left(\mathbf{d}^{e}_{pq}+\frac{\mathbf{d}^{nuc}}{N_{e}}\delta_{pq}\right)E_{pq}, \label{eq:dipole_definition} 
\end{equation}
with $\mathbf{d}^{e}$ being the electronic dipole operator and $\mathbf{d}^{nuc}$ the nuclear dipole moment of a system of $N_{e}$ electrons. The second quantization formalism for the electrons has been adopted in \cref{eq:Lenght_gauge} such that
\begin{equation}
\begin{split}
&E_{pq}=\sum_{\sigma}a^{\dagger}_{p\sigma}a_{q\sigma} \\ 
&e_{pqrs}=E_{pq}E_{rs}-\delta_{rq}E_{ps},\label{eq:Singlet}
\end{split}
\end{equation}
with $a^{\dagger}_{p\sigma}$ and $a_{p\sigma}$ respectively create and annihilate an electron in the orbital $p$ with spin $\sigma$. Finally, the electronic Hamiltonian in the Born-Oppenheimer approximation $H_{e}$ is defined as\cite{helgaker2013molecular}
\begin{equation}
H_{e}= \sum_{pq}h_{pq}E_{pq}+\frac{1}{2}\sum_{pqrs}g_{pqrs}e_{pqrs} , \label{eq:Standard_quantum}  
\end{equation}
where $h_{pq}$ and $g_{pqrs}$ are the one and two electron integrals. In the following, we label occupied orbitals in the HF reference with the letters $i,j,k...$ while the virtual orbitals are labeled $a,b,c..$. General orbital indices are labeled with $p,q,r,s$. 
In addition to the standard electronic terms, the strong coupling Hamiltonian in \cref{eq:Lenght_gauge} has three additional field-induced contributions, i.e. the purely photonic Hamiltonian $\omega b^{\dagger}b$, the bilinear light-matter term, $\lambda\sqrt{\frac{\omega}{2}} (    {\mathbf{d}}\cdot \pmb{\epsilon}) ( b^{\dagger}+b )$, explicitly correlating the field and the electrons, and finally the dipole self-energy (DSE) term, $\frac{\lambda^2}{2}  ( {\mathbf{d}} \cdot \pmb{\epsilon})^2$, needed to ensure that the Hamiltonian is bound from below.~\cite{rokaj2018light}

Rayleigh–Schrödinger (RS) perturbative schemes rely on a partition of the full Hamiltonian into a zeroth-order unperturbed Hamiltonian, $H^{(0)}$, whose eigenfunctions are known, and a perturbation.~\cite{lowdin1951note} The perturbation $V$ can be as complicated as needed to describe the physics of the overall system.
When $H^{(0)}$ is chosen to be the Fock operator from a mean-field theory, aimed at capturing the main physical properties of the system, we obtain the M{\o}ller-Plesset (MP) perturbation hierarchy.~\cite{moller1934note}
Accordingly, the perturbation $V$ is defined such that when added to the "solvable" zeroth-order Hamiltonian $H^{(0)}$, the full Hamiltonian is recovered
\begin{equation}
    V = H - H^{(0)} 
\end{equation}
and incorporates all the correlation effects of the many-body system.
The definition of an appropriate zeroth-order Hamiltonian is critical to ensure that perturbative methodologies provide a reliable description of the system, particularly if only a few orders in perturbation theory are considered.~\cite{helgaker2013molecular}

In this Section, we first review different choices for QED M{\o}ller-Plesset perturbation theory.~\cite{bauer2023perturbation} Then, we present the SC-QED-MP2 approach with the unperturbed Hamiltonian derived from the polaritonic mean field treatment of SC-QED-HF theory. Lastly, we review the LF-MP2 method with emphasis on the differences between this approach and the developed SC-QED-MP2 method.

\subsection{QED-MP2}
In the quantum electrodynamics Hartree-Fock method, the wave function is written as
\begin{equation}
\ket{\psi}= U_{\mathrm{QED-HF}} \ket{\mathrm{HF}} \otimes \ket{0} ,     
\end{equation}
which is composed of the bosonic vacuum $\ket{0}$ and an electronic Hartree-Fock Slater determinant, $\ket{\mathrm{HF}}$, where the low lying orbitals are occupied.
The coherent-state transformation
\begin{equation}\label{U qed-hf}
    U_{\mathrm{QED-HF}} = \exp\left(-z(b-b^{\dagger})\right)
\end{equation}
depends on the factor $z=\frac{\lambda}{\sqrt{2\omega}} \braket{\mathbf{d}\cdot\pmb{\epsilon}}$ which is updated throughout the SCF procedure together with the orbitals and 
\begin{equation}
    \braket{\mathbf{d}\cdot\pmb{\epsilon}} = \sum_{pq}\left(\mathbf{d}^{e}_{pq}+\frac{\mathbf{d}^{nuc}}{N_{e}}\delta_{pq}\right)D_{pq} ,
\end{equation}
where $D_{pq}=\bra{\mathrm{HF}}E_{pq}\ket{\mathrm{HF}}$ are the one-body density matrix elements.
It is convenient to change the quantum picture by applying the transformation to the light-matter Hamiltonian in \cref{eq:Lenght_gauge}
\begin{equation}\label{QED-HF Hamiltoninan}
    \begin{split}
    {H}_{\mathrm{QED-HF}} &= U^\dagger_{\mathrm{QED-HF}} \ H \ U_{\mathrm{QED-HF}} \\
    & = H_e + \omega  b^\dagger b + \frac{\lambda^2}{2}  ( (\mathbf{d}-\braket{\mathbf{d}}) \cdot \pmb{\epsilon})^2 \\
    & -\lambda\sqrt{\frac{\omega}{2}} ((\mathbf{d}-\braket{\mathbf{d}})\cdot \pmb{\epsilon}) ( b^{\dagger}+b ) ,
    \end{split}
\end{equation}
such that the origin-independence becomes explicit.
In this representation, the QED-HF wave function reads
\begin{equation}\label{reference wf}
    \ket{\mathrm{R}} = \ket{\mathrm{HF}}\otimes\ket{0} \equiv \ket{\mathrm{HF},0} .
\end{equation}
Bauer \textit{et al.}~\cite{bauer2023perturbation} proposed the QED-HF Fock operator plus the field energy $\omega b^{\dagger}b$ as the zeroth-order Hamiltonian for the QED-MP2 approach. Additionally, we highlight that the expectation value of the dipole squared ($\equiv \braket{\mathbf{d}\cdot\pmb{\epsilon}}^2$) should be included as well in the unperturbed Hamiltonian if not included in the perturbation $V$.~\cite{bauer2023perturbation}
The unperturbed Hamiltonian then reads  
\begin{equation}\label{zeroth H qedhf}
\begin{split}
    H^{(0)}_{\text{QED-HF}}&=\sum_{pq}F^{\text{QED-HF}}_{pq}E_{pq} \\
    &+\omega b^{\dagger}b+\frac{\lambda^{2}}{2}\braket{\mathbf{d}\cdot\pmb{\epsilon}}^2 ,
\end{split}
\end{equation}
where the Fock matrix elements are
\begin{equation}\label{fock qed-hf}
\begin{split}
     F^{\text{QED-HF}}_{pq}&=h_{pq}+\sum_{i}(2g_{pqii}-g_{piiq}) \\ & +\frac{\lambda^{2}}{2}\sum_{a}(\mathbf{d}\cdot\boldsymbol{\epsilon})_{pa}(\mathbf{d}\cdot\boldsymbol{\epsilon})_{aq} \\
     &-\frac{\lambda^{2}}{2}\sum_{i}(\mathbf{d}\cdot\boldsymbol{\epsilon})_{pi}(\mathbf{d}\cdot\boldsymbol{\epsilon})_{iq}.
\end{split}
\end{equation}
We point out that the coherent-state transformation does not change the electronic Hamiltonian. This will not be the case for the SC-QED-MP2 theory, and a correct transformation of the Fock matrix elements will be important to ensure orbital origin invariance.
For charged molecules, upon a shift $\mathbf{a}$ of the molecular system, the dipole integrals shift according to
\begin{equation} \label{dipole shift}
    ( \mathbf{d} \cdot \pmb{\epsilon} )_{pq} \rightarrow  ( \mathbf{d} \cdot \pmb{\epsilon} )_{pq} + \frac{Q_{tot}}{N_{e}} ( \mathbf{a} \cdot \pmb{\epsilon} ) \delta_{pq} ,
\end{equation}
where $Q_{tot}$ is the total system charge. As a consequence, the Fock matrix elements in \cref{fock qed-hf} are changed to
\begin{equation}
    \begin{split}
    F&^{\text{QED-HF}}_{pq} \rightarrow F^{\text{QED-HF}}_{pq} \\
    & + \frac{\lambda^2 (\delta_{pq,\text{vir}} - \delta_{pq,\text{occ}})}{2} 
    Q_{\text{tot}} (\mathbf{a} \cdot \boldsymbol{\epsilon}) (\mathbf{d} \cdot \boldsymbol{\epsilon})_{pq}
    \\
    & + \frac{\lambda^2 (\delta_{pq,\text{vir}} - \delta_{pq,\text{occ}})}{2}\frac{Q_{\text{tot}}^2 (\mathbf{a} \cdot \boldsymbol{\epsilon})^2 \delta_{pq}}{2} 
    \end{split}
\end{equation}
and this effect prevents the orbitals and their energies to be origin invariant.
Since the unperturbed Hamiltonian changes upon displacement of a charged molecular system, we expect an unphysical behavior of QED-MP2. 
Nonetheless, only rarely we do work with charged molecules and the problem can eventually be solved using the SC-QED-HF orbitals.
However, a more severe problem of the Fock operator in \cref{fock qed-hf} is the non size-intensivity. That is, for two systems A and B infinitely separated, the Fock operator is not equal to the sum of the two subsystems Fock operators
\begin{equation}
F^{\text{QED-HF}}_{AB}\neq F^{\text{QED-HF}}_{A}+F^{\text{QED-HF}}_{B}.    
\end{equation}
The orbital energies of the system $A$, instead, change if another system $B$ is added in the cavity regardless of the distance between $A$ and $B$. Upon insertion of molecule $B$, indeed, $F^{\text{QED-HF}}_{A}$ changes because the contribution from the nuclei of $B$ needs to be added in the dipole operator.
This might create problems when dealing with multi-component systems.
Nonetheless, we point out that the QED-HF method is size-extensive, i.e. the energy of a bipartite system where the subsystems A and B are far apart equals the sum of the individual subsystem energies. For this reason, QED-HF is unable to account for the cavity-induced non size-extensive effects.~\cite{haugland2021intermolecular}

The zeroth to second QED-MP energy corrections are given by the expressions
\begin{equation}
    E^{(0)}_\mathrm{QED-MP} = 2 \sum_i \epsilon^{\text{QED-HF}}_{i} + \lambda^{2}\braket{\mathbf{d}\cdot\pmb{\epsilon}}^2
 \end{equation}
\begin{equation}
    E^{(1)}_\mathrm{QED-MP}=-\sum_{ij} L^{\text{QED-HF}}_{iijj}
\end{equation}
\begin{equation}\label{eq:QED_MP2}
\begin{split}
    E^{(2)}_\mathrm{QED-MP}&=-\frac{1}{2}\sum_{aibj}\frac{L^{\text{QED-HF}}_{aibj}g^{\text{QED-HF}}_{aibj}}{\epsilon^{\text{QED-HF}}_{aibj}} \\
    & -\lambda^{2}\frac{\omega}{2}\sum_{ai}\frac{(\mathbf{d}\cdot\boldsymbol{\epsilon})^{2}_{ai}}{\epsilon^{\text{QED-HF}}_{ai}+\omega} .
\end{split}
\end{equation}
In the last equations, the redefined two electron integrals read
\begin{equation}\label{two electron int qed-hf}
    g^{\text{QED-HF}}_{pqrs}=g_{pqrs}+\lambda^{2}(\mathbf{d}\cdot\boldsymbol{\epsilon})_{pq}(\mathbf{d}\cdot\boldsymbol{\epsilon})_{rs} ,
\end{equation}
the integrals $L^{\text{QED-HF}}_{pqrs}$ are defined as
\begin{equation}\label{L qed-hf}
    L^{\text{QED-HF}}_{pqrs} = 2 g^{\text{QED-HF}}_{pqrs} - g^{\text{QED-HF}}_{psrq}
\end{equation}
and $\epsilon^{\text{QED-HF}}_{ai}$, $\epsilon^{\text{QED-HF}}_{aibj}$ showing in the denominators of the second order correction are
\begin{equation}\label{eps ai qedhf}
    \epsilon^{\text{QED-HF}}_{ai} = \epsilon^{\text{QED-HF}}_{a}-\epsilon^{\text{QED-HF}}_{i}  
\end{equation}
\begin{equation}\label{eps aibj qedhf}
    \epsilon^{\text{QED-HF}}_{aibj} = \epsilon^{\text{QED-HF}}_{ai} + \epsilon^{\text{QED-HF}}_{bj} ,
\end{equation}
where $\epsilon^{\text{QED-HF}}_{p}$ are the orbital energies obtained from the diagonalization of the QED-HF Fock matrix in \cref{fock qed-hf}.
Similarly to what happens with the standard MP2 scheme, the QED-HF energy is the sum of the zeroth and first-order energies.
We point out that neglecting the contribution $\propto\braket{\mathbf{d}\cdot\pmb{\epsilon}}^2$ in the zeroth-order Hamiltonian in \cref{zeroth H qedhf} would have led to the wrong QED-HF energy. Thus, the first non-vanishing correction to the QED-HF energy occurs in second-order of perturbation theory. The QED-MP2 correction in \cref{eq:QED_MP2} consists of two terms. The first term contains a contribution similar to the double excitations in the purely electronic MP2. However, it is important to highlight that the dipole contributions in $g^{\text{QED-HF}}_{pqrs}$ make the first term non size-extensive.
The second term in \cref{eq:QED_MP2}, instead, stems from the bilinear term of the Hamiltonian. It consists of contributions from single excitations in the electronic Hilbert space $\mathcal{H}_e$, coupled with single excitations in the photonic Hilbert space $\mathcal{H}_{ph}$ (double excitations in the polaritonic Hilbert space $\mathcal{H}_{pol} = \mathcal{H}_e \otimes \mathcal{H}_{ph}$). This term is size-extensive and partially cancels the DSE contribution to the QED-HF energy.~\cite{craig1998molecular}
Regarding size-intensivity, when changing the distance between two molecules that are already significantly far apart from each other, only the diagonal elements of the dipole matrix change significantly. Those elements enter the QED-MP2 energy correction in \cref{eq:QED_MP2} through the orbital energies in the denominator as
\begin{equation}
\begin{split}
    \epsilon^{\text{QED-HF}}_{p}&\propto \left(\frac{d^{NUC}_{A}}{N_{e}}+\frac{d^{NUC}_{B}}{N_{e}}\right)^{2}(\delta_{p,virt}-\delta_{p,occ}) \\
    &\propto (1+R_\text{AB})^{2}(\delta_{p,virt}-\delta_{p,occ})
\end{split}    
\end{equation}
where $R_\text{AB}$ is the distance between the two molecules.
Therefore easy to see that the QED-MP2 energy correction vanishes as $1/R_\text{AB}^{2}$.

\subsection{QED-(non-polaritonic HF)-MP2}
Even in the strong coupling regime, electron-electron correlation is dominating over the electron-photon one.
It is therefore reasonable to substitute the QED-HF Fock matrix in \cref{zeroth H qedhf} with the cavity free Hartree-Fock counterpart.
Proceeding with this choice, we end up with the QED non-polaritonic HF M{\o}ller-Plesset  perturbative scheme,~\cite{bauer2023perturbation} QED(np-HF)-MP2, with the zeroth-order Hamiltonian
\begin{equation}\label{eq:Unperturbed_HF}
\begin{split}
     H^{(0)}_{\text{QED-np-HF}}&=\sum_{pq}F^{\text{HF}}_{pq}E_{pq}+\omega b^{\dagger}b ,
\end{split}
\end{equation}
and the Fock matrix elements are 
\begin{equation}
F_{pq}^{HF}= h_{pq}+\sum_{j}(2g_{pqjj}-g_{pjjq}) .
\end{equation}
The reference wave function is again \cref{reference wf} and
the zeroth-order Hamiltonian in \cref{eq:Unperturbed_HF} does not account for any cavity effect on the electronic system. For this reason, as the coupling between light and matter increases, we expect the accuracy of QED(np-HF)-MP2 to diminish because the field effects become progressively more significant, and their inclusion in the zeroth-order (mean-field) Hamiltonian becomes important.
In this framework, the unperturbed Hamiltonian is both origin invariant and size-extensive, i.e. $H^{(0)}_{AB}=H^{(0)}_{A}+H^{(0)}_{B}$ for largely separated $A$ and $B$. Straightforwardly, when the two subsystems are infinitely far apart, the total energy reaches a plateau.

The zeroth to second order QED(np-HF)-MP energy corrections are given by the expressions
\begin{equation}
    E^{(0)}_\mathrm{QED(np-HF)-MP} = 2 \sum_i \epsilon^{\text{HF}}_{i} 
 \end{equation}
\begin{equation}\label{npHF mp1}
    E^{(1)}_{\text{QED(np-HF)-MP}}=-\sum_{ij}L_{iijj}+\lambda^{2}\sum_{ai}(\mathbf{d}\cdot\boldsymbol{\epsilon})^{2}_{ai}
\end{equation}
\begin{equation}\label{eq:npQED_MP2_HF} 
\begin{split}
E^{(2)}_{\text{QED(np-HF)-MP}}= &-\frac{1}{2}\sum_{aibj}\frac{L^{\text{QED-HF}}_{aibj}g^{\text{QED-HF}}_{aibj}}{\epsilon^{\text{HF}}_{aibj}} \\
& -\lambda^{2}\frac{\omega}{2}\sum_{ai}\frac{(\mathbf{d}\cdot\boldsymbol{\epsilon})^{2}_{ai}}{\epsilon^{\text{HF}}_{ai}+\omega} \\
& -\sum_{ai}\frac{\left(F^{\text{QED-HF}}_{ai}\right)^{2}}{\epsilon^{\text{HF}}_{ai}}  , 
\end{split}
\end{equation}
where the Fock matrix elements $F^{\text{QED-HF}}_{pq}$, the two electron integrals $g^{\text{QED-HF}}_{pqrs}$ and $L^{\text{QED-HF}}_{pqrs}$ are defined respectively in \cref{fock qed-hf,two electron int qed-hf,L qed-hf}. 
The definitions of $L_{pqrs}$, $\epsilon^{\text{HF}}_{ai}$ and $\epsilon^{\text{HF}}_{aibj}$ are analogous to the ones in \cref{L qed-hf,eps ai qedhf,eps aibj qedhf}.
The correction up to the first order (with \cref{npHF mp1}) of the zeroth-order energy gives the energy of the HF state in the cavity.
This energy is higher compared to its polaritonic counterpart since the orbitals are not optimized including the DSE contribution.
By comparing the second order energy correction in \cref{eq:npQED_MP2_HF} with the expression from QED-MP2 theory, we notice that the two first terms change in the denominators with the cavity-free HF orbital energies.
Moreover, an additional single electronic excitation contribution appears because the Brillouin theorem is not satisfied
\begin{equation}\label{kappa gradient}
   \bra{\mathrm{R}} \big[ {H}_{\mathrm{QED-HF}} , E^-_{ai}\big] \ket{\mathrm{R}} \neq 0  .
\end{equation}

\subsection{Strong coupling QED-MP2}
In the strong coupling quantum electrodynamics Hartree-Fock method, the wave function parametrization for the light-matter system is:
\begin{equation}
\ket{\psi_{\mathrm{SC}}}=\textrm{exp}\left(-\frac{\lambda}{\sqrt{2\omega}}\sum_{p}\eta_{p}\tilde{E}_{pp}(b-b^{\dagger})\right) \ket{\text{HF}}\otimes\ket{0},       
\label{eq:SC_QED_HF}
\end{equation}
where the tilde $\sim$ denotes integrals and operators in the basis that diagonalizes $(\mathbf{d}\cdot\boldsymbol{\epsilon})$:
\begin{equation}\label{mo to dipole}
    \sum_{rs}C_{rp}( \mathbf{d}\cdot\pmb{\epsilon})_{rs} C_{sq} = (\tilde{\mathbf{d}}\cdot\pmb{\epsilon})_{pp}\delta_{pq} ,
\end{equation}
where $\mathbf{C}$ is an orthonormal rotation matrix connecting molecular and dipole orbitals. The $\{\eta_p\}$ orbital specific coherent-state parameters in \cref{eq:SC_QED_HF} account for the field rearrangement to orbital excitations and are variationally optimized for the ground state through the SCF procedure.
Once again, it is convenient to change the quantum picture before partitioning the Hamiltonian into $H^{(0)}$ and $V$.
We apply the SC-transformation
\begin{equation}\label{U SC}
    U_{\mathrm{SC}} = \exp(-\frac{\lambda}{\sqrt{2\omega}} \sum_p \eta_p \Tilde{E}_{pp} ( b - b^\dagger )) 
\end{equation}
to the Pauli-Fierz Hamiltonian  in \cref{eq:Lenght_gauge}
\begin{equation}\label{SC Hamiltoninan}
    \begin{split}
    {H}_{\mathrm{SC}} &= U^\dagger_{\mathrm{SC}}  \ H \ U_{\mathrm{SC}} \\
    & = \sum_{pq} \tilde{h}_{pq}^{\text{SC}} Y_{pq}\tilde{E}_{pq}
    + \frac{1}{2}\sum_{pqrs}\tilde{g}_{pqrs}^{\text{SC}}Y_{pqrs}\tilde{e}_{pqrs} \\
    & + \omega b^\dagger b - \lambda\sqrt{\frac{\omega}{2}}\sum_p(\tilde{\mathbf{d}}\cdot\pmb{\epsilon})_{pp}^\eta \tilde{E}_{pp} (b + b^\dagger) ,
    \end{split}
\end{equation}
where we defined the $\eta$-shifed dipole integrals as 
\begin{equation}
(\tilde{\mathbf{d}}\cdot\pmb{\epsilon})^\eta_{pq} = (\tilde{\mathbf{d}}\cdot\pmb{\epsilon})_{pq} - \eta_p \delta_{pq}.
\end{equation}
The reference wave function $\ket{\mathrm{R}}$ is again defined in \cref{reference wf}.
The redefined SC one and two electron integrals entering in \cref{SC Hamiltoninan} read as
\begin{equation}\label{h_SC}
    \tilde{h}_{pq}^{\mathrm{SC}} = \tilde{h}_{pq} + \frac{\lambda^2}{2}((\tilde{\mathbf{d}}\cdot\pmb{\epsilon})_{pp}^\eta)^2 \delta_{pq} 
\end{equation}
\begin{equation}\label{g_SC}
    \tilde{g}_{pqrs}^{\mathrm{SC}} = \tilde{g}_{pqrs} + {\lambda^2}(\tilde{\mathbf{d}}\cdot\pmb{\epsilon})_{pp}^\eta(\tilde{\mathbf{d}}\cdot\pmb{\epsilon})_{rr}^\eta\delta_{pq}\delta_{rs} ,
\end{equation}
while, the $Y_{pq}$ and $Y_{pqrs}$ photonic operators are defined as follows
\begin{equation}
    Y_{pq} = \exp(\frac{\lambda}{\sqrt{2\omega}}(\eta_p-\eta_q)(b-b^\dagger)) 
\end{equation}
\begin{equation}
    Y_{pqrs} = \exp(\frac{\lambda}{\sqrt{2\omega}}(\eta_p-\eta_q+\eta_r-\eta_s)(b-b^\dagger)) .
\end{equation}
In line with the previous approaches, we define the zeroth-order Hamiltonian as 
\begin{equation}\label{H_zero SC}
\begin{split}
H^{(0)}_{\text{SC}}&=\sum_{pq}{F}^{\text{SC}}_{pq}{E}_{pq}+\omega b^{\dagger}b .
\end{split}
\end{equation}
For this approach, there is no need to add the expectation value of the dipole squared contribution because the SC-transformation in \cref{U SC} shifts the bosonic operators linearly with the $\{\eta_p\}$ parameters.
These contributions are already reabsorbed in the SC-redefined integrals in \cref{h_SC,g_SC}.
This is not the case for the previous methods where the shift brought by the QED-HF transformation in \cref{U qed-hf} is $\propto \braket{\mathbf{d}\cdot\pmb{\epsilon}}$.
Furthermore, as mentioned earlier, the effect of the SC-transformation in \cref{U SC} on the Fock matrix elements
\begin{equation}\label{SC fock}
    \Tilde{F}^{\text{SC}}_{pq} = \tilde{h}_{pq}^{\text{SC}} Q_{pq} + \frac{1}{2}\sum_{rs}\tilde{L}_{pqrs}^{\text{SC}}  Q_{pqrs} \Tilde{D}_{rs}
\end{equation}
is non-trivial due to the introduction of the gaussian factors
\begin{equation}
\begin{split}
    Q_{pq} & = \braket{Y_{pq}}_0 \\
    & = \exp(-\frac{\lambda^2}{4\omega}(\eta_p - \eta_q)^2)
\end{split} 
\end{equation}
\begin{equation}
\begin{split}
    Q_{pqrs} &= \braket{Y_{pqrs}}_0 \\
    & = \exp(-\frac{\lambda^2}{4\omega}(\eta_p - \eta_q +\eta_r - \eta_s +)^2) ,
\end{split}
\end{equation}
that carry the $\omega$-correlation captured at the mean field level.
The integrals $\tilde{L}_{pqrs}^{\text{SC}}$ read
\begin{equation}
\tilde{L}_{pqrs}^{\text{SC}} = 2 \tilde{g}_{pqrs}^{\text{SC}} - \tilde{g}_{psrq}^{\text{SC}} ,
\end{equation}
while the Fock matrix elements in \cref{H_zero SC,SC fock} are connected by
\begin{equation}
    \sum_{rs}C_{rp} F_{rs} C_{sq} = \tilde{F}_{pq} ,
\end{equation}
where the canonical to dipole basis transformation $\mathbf{C}$ is defined in \cref{mo to dipole}.
The SC-QED-HF Fock is fully origin invariant as any displacement of the system (i.e. a change of the diagonal elements of $(\tilde{\mathbf{d}}\cdot\boldsymbol{\epsilon})$) is readily reabsorbed by an appropriate change of the optimal $\{\eta_p\}$ parameters.
Moreover, the Fock matrix elements in \cref{SC fock} are also size-intensive as demonstrated in Ref.~\citenum{riso2022molecular}.
The orbital specific coherent-state transformation in \cref{eq:SC_QED_HF} inherently introduces correlation between electrons and photons, with significant implications on the perturbative energy corrections.
The zeroth and first order energy corrections read
\begin{equation}
    E^{(0)}_\mathrm{SC-QED-MP} = 2 \sum_i \epsilon^{\text{SC}}_{i} 
 \end{equation}
\begin{equation}
    E^{(1)}_{\text{SC-QED-MP}} =  -\sum_{ij}(2g^{\text{SC-Q}}_{iijj}-g^{\text{SC-Q}}_{ijji})
\end{equation}
and, once again, their sum leads to the SC-QED-HF energy. 
With $\epsilon^{\text{SC}}_{p}$ we refer to the orbital energies obtained from the diagonalization of the SC Fock matrix in \cref{SC fock}. The two electron integrals in the first order correction obtained from the dipole to canonical basis transformation of the integrals in \cref{g_SC} using the same $\mathbf{C}$ matrix used to change basis of the Fock matrix
\begin{equation}
    g^{\text{SC-Q}}_{pqrs}= \sum_{tuvz}C_{pt}C_{rv} \ \tilde{g}_{tuvz}^{\mathrm{SC}} \  Q_{tuvz} \ C_{qu} C_{sz} ,
\end{equation}
with inclusion of the Gaussian factors $Q_{pqrs}$.

The SC-QED M{\o}ller-Plesset second order energy correction is given by
\begin{equation}
E^{(2)}_\mathrm{SC-QED-MP} = 
-\frac{1}{2}\sum^{\infty}_{n=0}\sum_{aibj}\frac{\bra{\text{R}}H_{\mathrm{SC}}\ket{^{ab}_{ij},n}^{2}}{n\omega+\epsilon_{aibj}^{\text{SC}}}
-\sum^{\infty}_{n=1}\sum_{ai}\frac{\bra{\text{R}}H_{\mathrm{SC}}\ket{^{a}_{i},n}^{2}}{n\omega+\epsilon_{ai}^{\text{SC}}} 
- \sum^{\infty}_{n=2}\frac{\bra{\text{R}}H_{\mathrm{SC}}\ket{\text{HF},n}^{2}}{n\omega} ,
\label{eq:SC_QED_MP2}
\end{equation}
where the index $n$ refers to the number of photons in the respective excited determinant and
\begin{equation}
    \epsilon^{\text{SC}}_{ai} = \epsilon^{\text{SC}}_{a}-\epsilon^{\text{SC}}_{i}  
\end{equation}
\begin{equation}
    \epsilon^{\text{SC}}_{aibj} = \epsilon^{\text{SC}}_{ai} + \epsilon^{\text{SC}}_{bj} .
\end{equation}
For a more detailed and explicit derivation of \cref{eq:SC_QED_MP2} we refer the reader to the  Supporting Information where we also derive the more general perturbation theory for multiple sets of bosons and modes coupled to an electronic Hamiltonian.
We highlight that the Hamiltonian now connects $\ket{\text{HF},0}$ and determinants that include more than one photon. The excitations contributing to the second-order energy correction can be divided into three classes:
\begin{enumerate}
    \item Double excitations in the electronic reference with an arbitrary photon number $\ket{^{ab}_{ij},n \ge 0}$. The contribution for $n=0$ is equivalent to the first term in eq \ref{eq:QED_MP2} while all the other terms are SC-QED-MP2 specific;
    \item Single excitations in the electronic reference with a photon number larger than zero $\ket{^{a}_{i},n\ge1}$. The contribution for $n=0$ is zero because of the SC-QED-HF Brillouin condition in the orbital part
    \begin{equation}
        \bra{\mathrm{R}} \big[ {H}_{\mathrm{SC}} , E^-_{ai}\big] \ket{\mathrm{R}} = 0 .
    \end{equation}
    The $\ket{^{a}_{i},1}$ term incorporates the light-matter bilinear contribution;
    \item Excitations in the field part only $\ket{\text{HF},n\ge2}$. These terms are specific to SC-QED-MP2.
    Their presence demonstrates that in the SC-QED-MP2 scheme photons and electrons are treated on an equal footing.
     The contribution for $n=1$ is null because of the SC-QED-HF Brillouin condition in the photonic part
    \begin{equation}
        \bra{\mathrm{R}} \big[ {H}_{\mathrm{SC}} , \Tilde{E}_{pp}(b-b^\dagger)\big] \ket{\mathrm{R}} = 0 .
    \end{equation}
\end{enumerate} 

\subsection{Lang-Firsov MP2}
Recently, a M{\o}ller-Plesset perturbative approach up to the fourth order and based on a wave function similar to the SC-QED-HF one has been published.~\cite{cui2024variational} The LF-HF wave function parametrization reads
\begin{equation}
\ket{\psi_{\mathrm{LF}}}= U_{\mathrm{LF}} \ U_{\mathrm{CS}} \ \ket{\text{HF}}\otimes\ket{0} ,  
\label{eq:LF_HF}
\end{equation}
where the transformations are given by
\begin{equation}\label{U_LF}
    U_{\mathrm{LF}} = \prod_\alpha\textrm{exp}\left(-\frac{\lambda}{\sqrt{2\omega}}\sum_{p}\eta_{p}^\alpha\bar{E}_{pp}(b_\alpha-b^{\dagger}_\alpha)\right)
\end{equation}
\begin{equation}
    U_{\mathrm{CS}} = \prod_\alpha\textrm{exp}\left(-z_\alpha(b_\alpha-b^{\dagger}_\alpha)\right) .
\end{equation}
Here $\alpha$ denotes cavity modes, while $\{\eta_{p}^\alpha\}$ and $\{z_\alpha\}$ are variational parameters.
The difference with respect to SC-QED-MP2 is the basis choice. 
In \cref{U_LF}, the $\bar{E}_{pp}$ operator is in the orthogonalized Löwdin basis, which differs from the dipole basis that diagonalizes $(\mathbf{d}\cdot\boldsymbol{\epsilon})$.
In particular, the atomic basis functions are orthogonalized by means of the $\mathbf{S}^{-\frac{1}{2}}$ matrix, where $\mathbf{S}$ is the overlap matrix between the AOs.

Furthermore, the total transformation in \cref{eq:LF_HF} is redundant.
In fact, the two transformations commute and $U_{\mathrm{CS}}$ can be reabsorbed in $U_{\mathrm{LF}}$ by an appropriate shift of the $\{\eta_{p}^\alpha\}$ parameters. For this reason, in the remaining of this Section we neglect the $\{z_\alpha\}$ parameters.
Using a generic basis seems advantageous for developing a multi-mode \textit{ab initio} polaritonic theory.
However, there are physical reasons that keep us relying on the dipole basis.
Not only in the dipole basis is the wave function exact in the infinite coupling limit, but not using it may lead to non size-intensive molecular orbitals.
In fact, from a preliminary theoretical investigation, the single-mode DSE contribution to the Fock matrix elements is
\begin{equation}\label{fock LF}
\begin{split}
    \bar{F}_{pq}^{\mathrm{DSE}} & = \frac{\lambda^2}{2} \sum_r (\bar{\mathbf{d}}\cdot\pmb{\epsilon})^\eta_{pr}  (\bar{\mathbf{d}}\cdot\pmb{\epsilon})^\eta_{rq} Q_{pq} \\
    & +  \lambda^2 (\bar{\mathbf{d}}\cdot\pmb{\epsilon})^\eta_{pq} \sum_{rs} (\bar{\mathbf{d}}\cdot\pmb{\epsilon})^\eta_{rs} Q_{pqrs} \bar{D}_{rs} \\
    & - \frac{\lambda^2}{2} \sum_{rs} (\bar{\mathbf{d}}\cdot\pmb{\epsilon})^\eta_{ps}(\bar{\mathbf{d}}\cdot\pmb{\epsilon})^\eta_{rq} Q_{pqrs} \bar{D}_{rs} ,
\end{split}
\end{equation}
For a bipartite system where the subsystems A and B are far apart, the $r$ and $s$ summation in the Coulomb term (second line in \cref{fock LF}) is over both A and B orbitals. This implies that the Fock matrix, as in QED-MP2, is not size-intensive because of the contributions from system B to the $\bar{F}_{p_Aq_A}^{\mathrm{DSE}}$ elements.
The theoretical argument extends straightforwardly to the multi-mode case.

\section{3. Results and discussions}
In this Section, we assess the performance of the QED-MP2, QED(np-HF)-MP2, LF-MP2 and the developed SC-QED-MP2 methods.
We focus on cavity coupling and frequency dispersions, intermolecular potential energy curves for different kinds of interactions and lastly orientational effects of the polarization vector with respect to the molecular system.
We use the QED-CCSD \cite{haugland2020coupled} as the benchmark for the perturbative results.
Specifically, this coupled cluster theory is obtained using QED-HF as a reference wave function and is expected to capture more correlation energy than SC-QED-MP2 due to the inclusion of unlinked excitations entering the many-body exponential cluster operator.
For this reason, although the SC-QED-MP2 method is based on a different reference wave function, the comparison is justified as we focus on electron-photon correlation effects.
The SC version of QED-CC is currently under development and we expect it to capture more correlation than its QED-HF counterpart.
The QED-MP2, QED(np-HF)-MP2 and SC-QED-MP2 methods have been implemented in a development version of the $e^\mathcal{T}$ program.~\cite{folkestad2020t} The LF-MP2 calculations have been performed using the \textsc{Polar} program,~\cite{cui_polar} which is supported with some routines from the \textsc{PySCF} package.~\cite{sun2020recent} All calculations have been performed using the aug-cc-pVDZ basis set.~\cite{pritchard2019a,dunning1989a}
The molecular structures have been optimized using the ORCA software\cite{neese2022software} with DFT-B3LYP functional and using a def2-SVP basis set.
We point out that to compare the  M{\o}ller-Plesset perturbative methods with the QED-CCSD theory, one should focus on the qualitative trend of the curves rather than total energies.
Perturbative methods, as well as coupled cluster approaches, are non-variational and lower energies are not necessarily indicative of better performance.
For the results obtained with the SC-QED-MP2 approach, the infinite summation in the photonic space in \cref{eq:SC_QED_MP2} is truncated when the contributions are smaller than $10^{-12} \ a.u.$ In the LF-MP2 calculations the truncation is made after considering 17 photons in the photonic space, which is more than enough to converge the results.

\subsection{Cavity coupling and frequency dispersions}
\begin{figure}[!ht]
    \centering
    \includegraphics[width=0.8\textwidth]{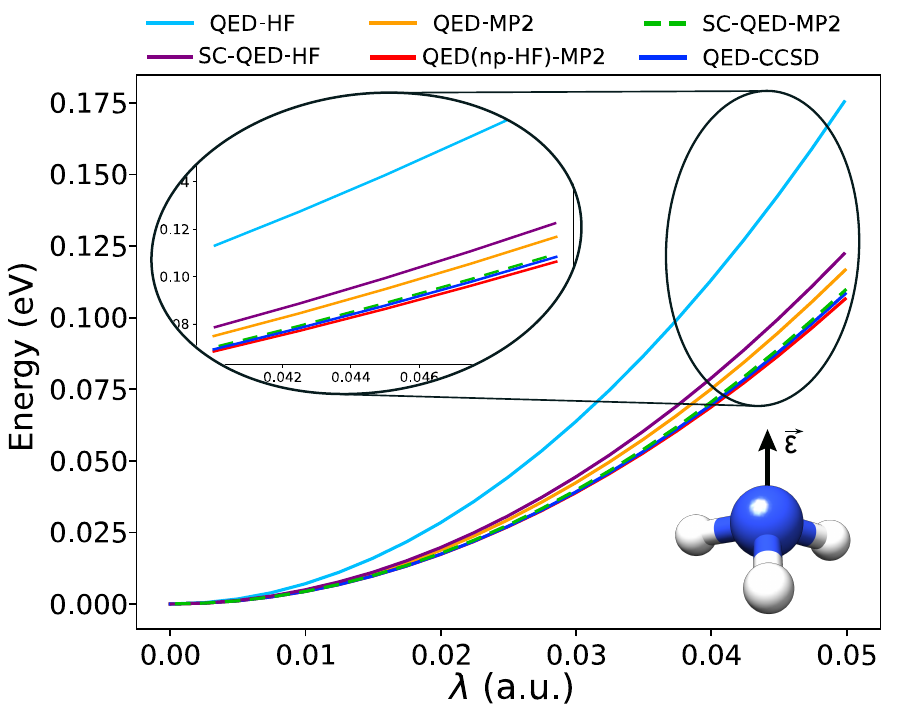}
    \caption{Coupling dispersions for an ammonia molecule. The cavity frequency is set to $\omega =$ \qtylist{8.16}{\electronvolt}, while the polarization $\pmb{\epsilon}$ is along the $\mathrm{C}_3$ axis. For realistic coupling values, $\lambda\leq$ \qtylist{0.05}{\atomicunit}, all the methods show the same increasing trend. For SC-QED-MP2, the inclusion of electron-photon correlation at the mean-field level becomes important for larger couplings.}
    \label{fig:Lambda_dispersion}
\end{figure}
In Figure \ref{fig:Lambda_dispersion}, we show the energy dispersions of an ammonia molecule in an optical cavity with a frequency of $\omega =$ \qtylist{8.16}{\electronvolt} as a function of the light-matter coupling parameter $\lambda$. The cavity polarization is aligned along the $\mathrm{C}_3$ symmetry axis of the molecule.
The $\lambda=$ \qtylist{0}{\atomicunit} energy is set to zero for all the methods. All the presented methodologies capture the qualitative effect of the field, i.e. the energy increases with increasing light-matter coupling.
The mean-field approaches, i.e. the QED-HF and SC-QED-HF, overestimate this trend, which is decreased by including more electron-photon correlation. However we observe that the SC-QED-HF approach performs better by capturing some electron-photon correlation already at the mean-field level.~\cite{riso2022molecular,el2024toward}
Within the M{\o}ller-Plesset methods, QED-MP2 performs worse than the others, overestimating the trend of QED-CCSD and lying close to SC-QED-HF at higher couplings.
This is not surprising considering the ill-defined molecular orbitals of the reference QED-HF.
The QED(npHF)-MP2 and SC-QED-MP2 methods perform well, following for all coupling values the QED-CCSD trend.
However, as discussed in Section 3, we expect that for very strong values of the light-matter coupling the QED(np-HF)-MP2 should exhibit a decrease in accuracy.
This effect is displayed in the zoom panel on the left where, at higher couplings, the SC-QED-MP2 becomes the most accurate methodology.
This observation is in agreement with the fact that the reference SC-QED-HF wave function becomes exact in the infinite coupling limit.
However, we point out that realistic $\lambda$ values for strongly coupled systems are smaller than \qtylist{0.05}{\atomicunit} (corresponding to a quantization volume of around 1 nm$^{3}$).

\begin{figure}[!ht]
    \centering
    \includegraphics[width=0.8\textwidth]{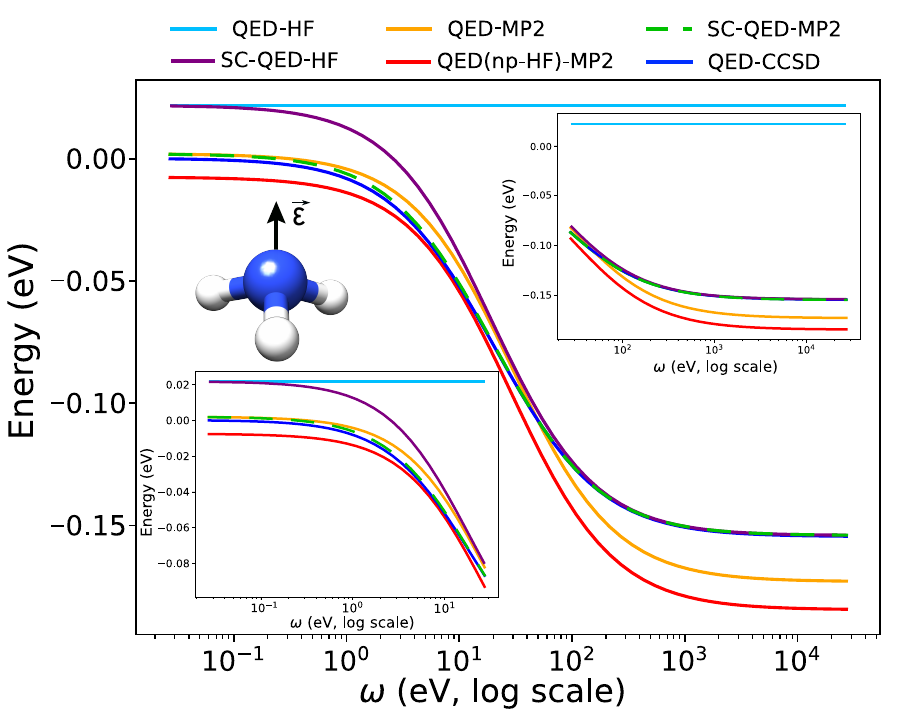}
    \caption{Frequency dispersions for an ammonia molecule. The cavity light-matter coupling is set to $\lambda =$ \qtylist{0.05}{\atomicunit}, while the polarization $\pmb{\epsilon}$ is along the $\mathrm{C}_3$ axis. The SC-QED-MP2 approach reproduces well the QED-CCSD trend for the whole range of $\omega$.}
    \label{fig:freq}
\end{figure}
In Figure \ref{fig:freq}, we plot the cavity frequency energy dispersions for the same system with the light-matter coupling set to $\lambda=$ \qtylist{0.05}{\atomicunit}
The offset is chosen in order to unbias the comparisons with respect to the electron-electron correlation.
To this end, the QED-HF and SC-QED-HF curves are shifted by the electron-electron correlation captured by CCSD outside the cavity.
On the other hand, for QED-MP2, QED(np-HF)-MP2 and SC-QED-MP2 we shift the curves by the difference of the electron-electron correlation between MP2 and CCSD outside the cavity.
Finally, the zero energy point is equal to the QED-CCSD results at low frequencies.
One of the main strengths of SC-QED-HF is its ability to exhibit a qualitatively correct frequency dispersion of the energy, even at the mean-field level.~\cite{riso2022molecular,el2024toward}
The SC-QED-HF curve matches the QED-CCSD dispersion at extremely high frequencies.
In fact, in that range of $\omega$, light and matter are effectively decoupled.
At small cavity frequencies, the SC-QED-HF curve overestimates the QED-CCSD results because of the additional electron-photon correlation captured by coupled cluster.
In contrast, the QED-HF method does not display any correlation at all and the energy remains constant.
All the M{\o}ller-Plesset methodologies display the correct dispersion trend.
However, the QED-MP2 and QED(np-HF)-MP2 methods perform worse compared to SC-QED-MP2, which matches the QED-CCSD results for the whole frequency range.
We claim that this happens because the electron-photon correlation is captured only perturbatively by QED-MP2 and QED(np-HF)-MP2. The SC-QED-MP2, on the other hand, relies on a cavity-consistent set of molecular orbitals and the electron-photon correlation is already naturally included in the zeroth-order Hamiltonian.

\subsection{Intermolecular interactions}
\begin{figure}[!ht]
    \centering
    \includegraphics[width=\textwidth]{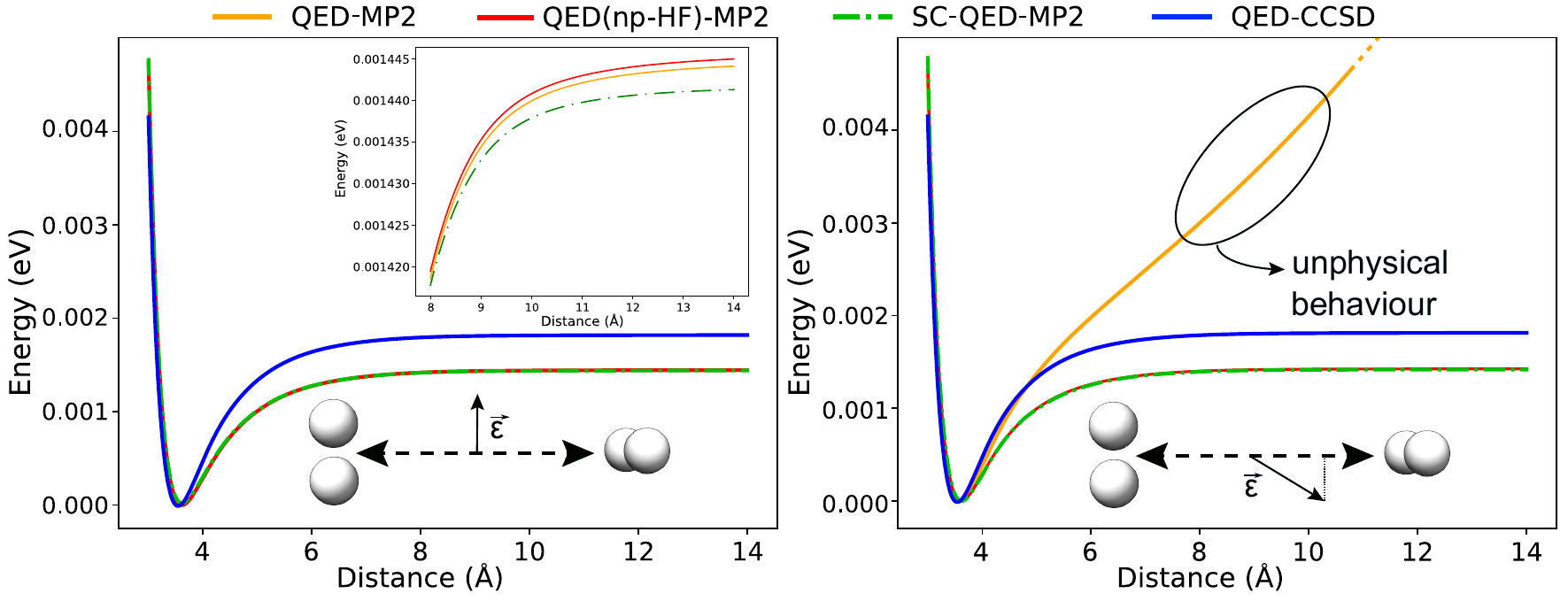}
    \caption{Dissociation curves for two $\mathrm{H}_2$ molecules in an optical cavity with frequency and light-matter coupling set to $\omega=$ \qtylist{27.2}{\electronvolt} and $\lambda=$ \qtylist{0.01}{\atomicunit} On the left the polarization $\pmb{\epsilon}$ is orthogonal to the displacement direction, while on the right it has a component $1/\sqrt{3}$. When the polarization has a component along the displacement direction the QED-MP2 method displays an unphysical behavior in the long-range regime.} 
    \label{fig:H2}
\end{figure}
Long-range effects become significant when intermolecular interactions are considered, for example in the case of the van der Waals interaction between two hydrogen molecules shown in Figure \ref{fig:H2}.
In particular, we plot the dissociation potential energy curves of the M{\o}ller-Plesset approaches.
All the curves are shifted such that the minima are set to zero. The cavity frequency and the light-matter coupling are set to $\omega=$ \qtylist{27.2}{\electronvolt} and $\lambda=$ \qtylist{0.01}{\atomicunit}
On the left we show the results for a polarization along to the $z$ axis (orthogonal to the displacement direction).
On the right, instead, we plot the results for a field polarization of $\boldsymbol{\epsilon}=\left(\frac{1}{\sqrt{3}},\frac{1}{\sqrt{3}},-\frac{1}{\sqrt{3}}\right)$.
For the first polarization we notice that all the methods provide a good description around the equilibrium distance.
The M{\o}ller-Plesset methods overestimate the attractive part of the potential $\propto-1/r^6$.
From the second polarization on the right, the QED(npHF)-MP2 and SC-QED-MP2 perform in a similar way, while the QED-MP2 displays an unphysical behavior in the long-range regime.
For symmetry arguments, this behavior has to be due to the polarization component along the displacement direction.
We stress that the system is not charged.
For this reason the issue does not emerge from the origin dependent molecular orbitals for charged systems.
The problem stems from the non size-intensivity of the zeroth-order Hamiltonian which is based on ill-defined orbital energies for separated systems.

\begin{figure}[!ht]
    \centering
    \includegraphics[width=\textwidth]{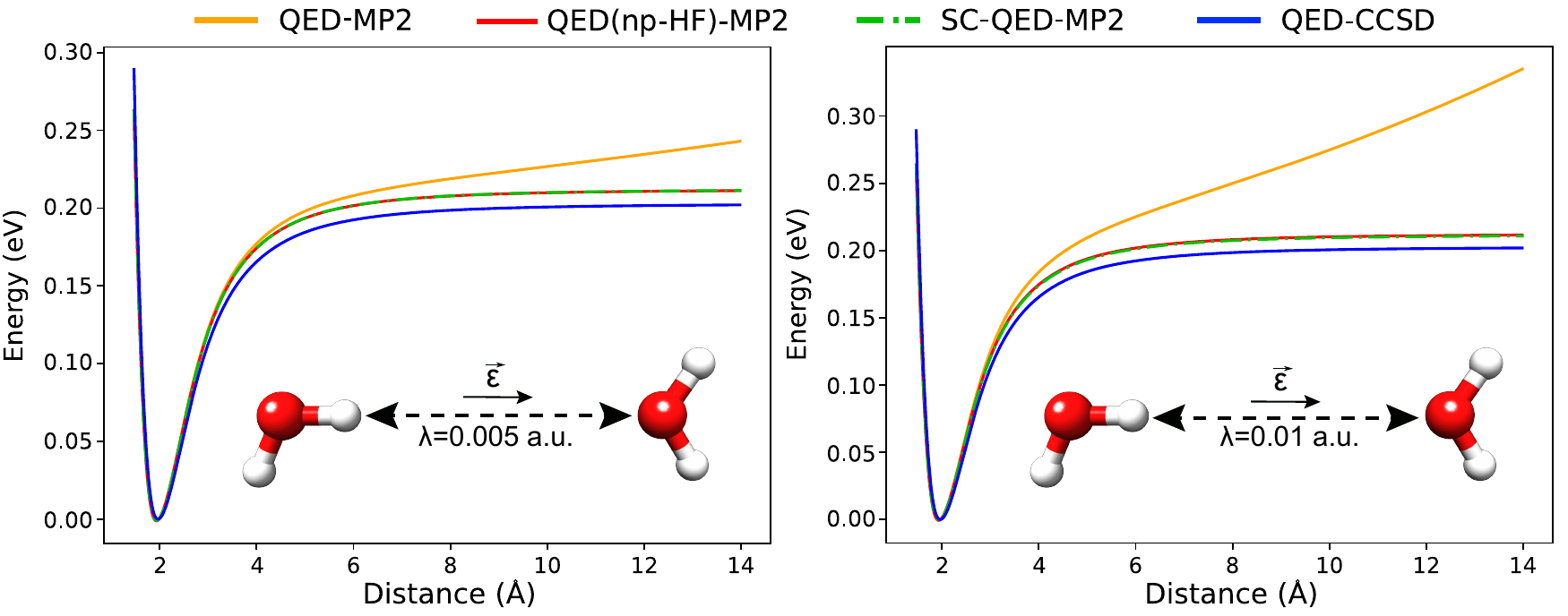}
    \caption{Dissociation curves for two water molecules in a hydrogen bonding geometry inside a cavity. The frequency is set to $\omega=$ \qtylist{8.16}{\electronvolt}, while the polarization $\pmb{\epsilon}$ is along to the displacement direction. The light-matter coupling is set to $\lambda=$ \qtylist{0.005}{\atomicunit} on the left and $\lambda=$ \qtylist{0.01}{\atomicunit} on the right. The unphysical behavior displayed by QED-MP2 is enhanced at higher couplings.}
    \label{hydrogen_bond}
\end{figure}
In Figure \ref{hydrogen_bond}, we investigate the behavior of the M{\o}ller-Plesset approaches for a hydrogen-bonded geometry of the water dimer. The leading term of the intermolecular interaction is given by the dipole-dipole interaction, however, the charge-transfer component along the hydrogen bridge is known to be non-negligible.~\cite{ronca2014quantitative}
The cavity frequency is set to $\omega=$ \qtylist{8.16}{\electronvolt}, while the light-matter coupling is set to $\lambda=$ \qtylist{0.005}{\atomicunit} in the left plot and $\lambda =$ \qtylist{0.01}{\atomicunit} in the right one. For both the plots the polarization vector is along the $y$ axis (the displacement direction).
The QED(np-HF)-MP2 and SC-QED-MP2 reproduce qualitatively well the potential curve but underestimate the binding energy, contrary to what is observed for the van der Waals interaction in Figure \ref{fig:H2}. 
With the polarization along the displacement direction, the QED-MP2 approach displays again an unphysical behavior due to the ill-defined molecular orbitals of the reference QED-HF.
Comparing the plots at different couplings we can see that the issue is enhanced with increasing the light-matter coupling $\lambda$.
This observation is consistent with the $\lambda^2$-scaling of the non size-intensive terms of the Fock matrix elements in \cref{fock qed-hf}.

\begin{figure}[!ht]
    \centering
    \includegraphics[width=\textwidth]{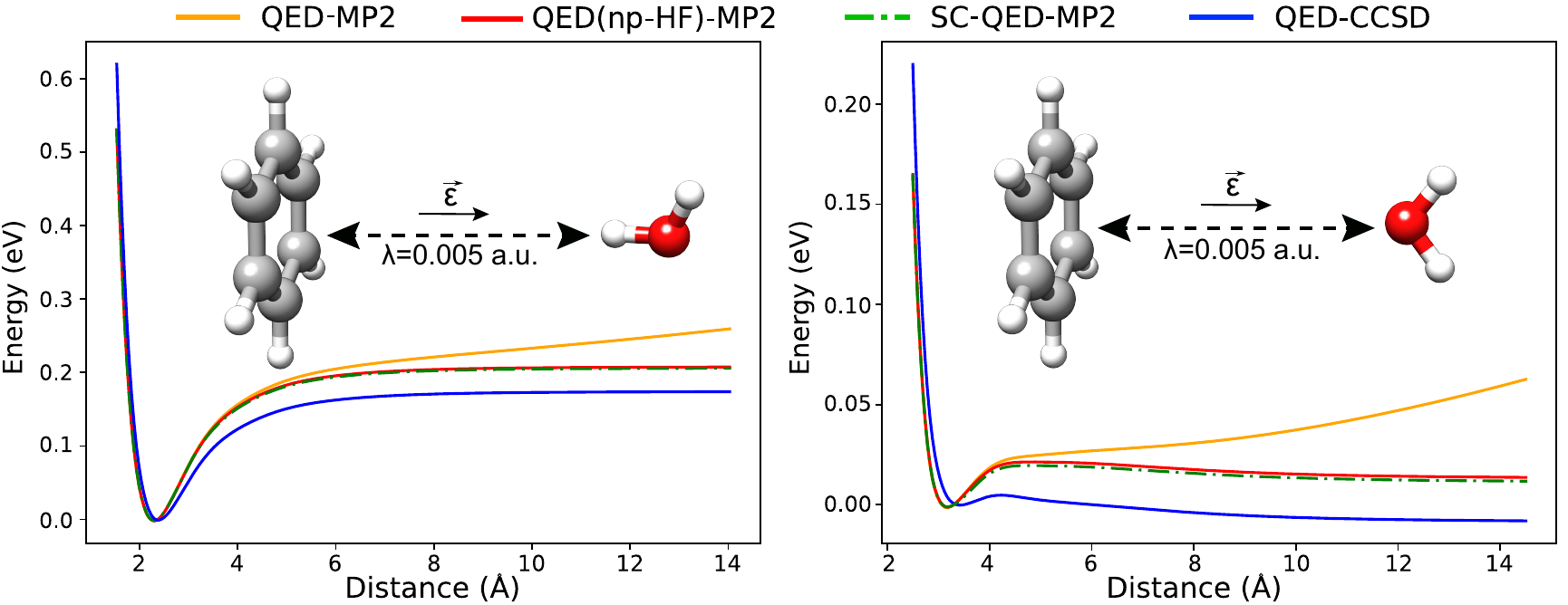}
    \caption{Dissociation curves inside a cavity for a benzene and a water molecule in two different geometries. For both cases the frequency is set to $\omega=$ \qtylist{2.27}{\electronvolt}, while light-matter the coupling is set to $\lambda=$ \qtylist{0.005}{\atomicunit} The polarization $\pmb{\epsilon}$ is along to the displacement direction. The unphysical behavior displayed by QED-MP2 changes an unbounded intermolecular interaction into a bounded one (see plot on right).}
    \label{dip_induced_dip}
\end{figure}
In Figure \ref{dip_induced_dip}, we show the behavior of the perturbative approaches for a dipole-induced dipole system composed by a benzene and a water molecule.
The polarization vector $\pmb{\epsilon}$ is again along the displacement direction, while the light-matter coupling and cavity frequency are set to $\lambda =$ \qtylist{0.005}{\atomicunit} and $\omega =$ \qtylist{2.27}{\electronvolt}.
On the left, the system is bonded because one of the water hydrogens points toward the benzene.
On the right, instead, the system shows a metastable minima because the oxygen of the water molecule is pointing toward the ring.
This is due to the repulsive interaction between the $\pi$ electrons cloud of the benzene and the lone pairs on the oxygen atom.
Yet again, the trend of the M{\o}ller-Plesset approaches is the same, with the QED-MP2 displaying an unphysical behavior. However, for the metastable sytem on the right, this issue is even more troublesome because a physically unbounded system is turned into a bounded one.

\begin{figure}[H]
    \centering
    \includegraphics[width=0.7\textwidth]{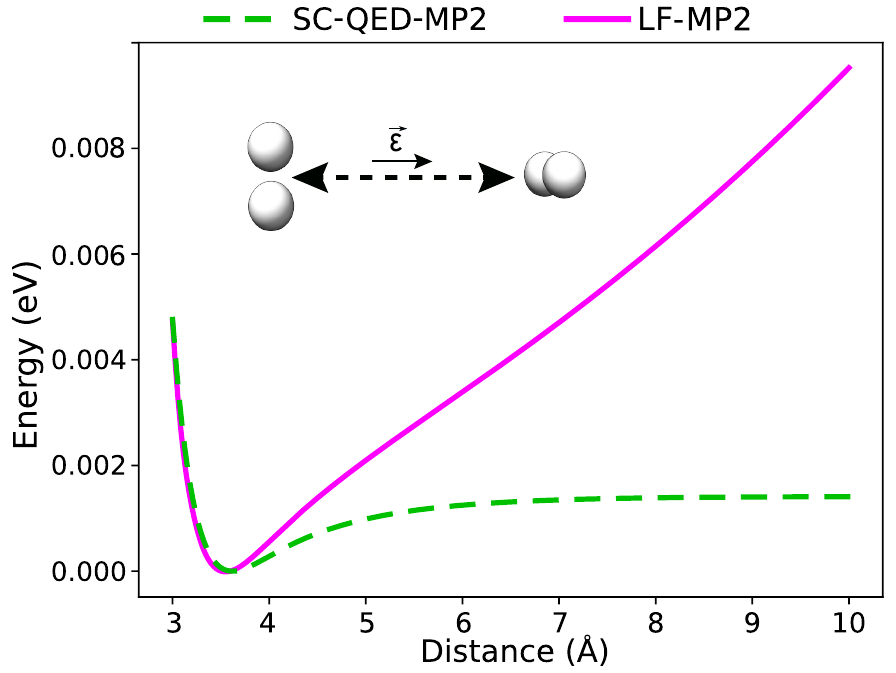}
    \caption{Comparison between SC-QED-MP2 and LF-MP2 dissociation curves for two $\mathrm{H}_2$ molecules in an optical cavity with frequency and light-matter coupling set to $\omega=$ \qtylist{2.72}{\electronvolt} and $\lambda=$ \qtylist{0.01}{\atomicunit} The polarization $\pmb{\epsilon}$ is along the displacement direction. The LF-MP2 method displays the same unphysical behavior as QED-MP2.} 
    \label{sc VS lf}
\end{figure}

From the results shown, is clear that SC-QED-MP2 and QED(npHF)-MP2 are well behaved and capture the different kinds of intermolecular interactions.
Using a wave function parametrization similar to SC-QED-HF, but in another basis, is not sufficient to obtain the similar accuracy.
In Figure \ref{sc VS lf}, we show the comparison between SC-QED-MP2 and LF-MP2 for the two hydrogen van der Waals system.
The cavity frequency and the light-matter coupling are set to $\omega=$ \qtylist{2.72}{\electronvolt} and $\lambda=$ \qtylist{0.01}{\atomicunit}, while the polarization vector $\pmb{\epsilon}$ is along the displacement direction.
The LF-MP2 curve behaves like the one for QED-MP2 by displaying the same unphysical diverging trend in the long-range regime.
This behavior could be due to the non size-intensivity of the Fock matrix in a basis that differs from the dipole basis as shown in \cref{fock LF}. An alternative explanation for this behavior has an algorithmic nature. The existence of a bifurcation point in the parameter space that is not accurately treated in the optimization procedure can lead to the convergence to another solution different from the ground state.
Further investigations are necessary, but these observations suggest that the choice of basis is crucial in \textit{ab initio} polaritonic models.

\subsection{Polarization orientational effects}
\begin{figure}[H]
    \centering
    \includegraphics[width=\textwidth]{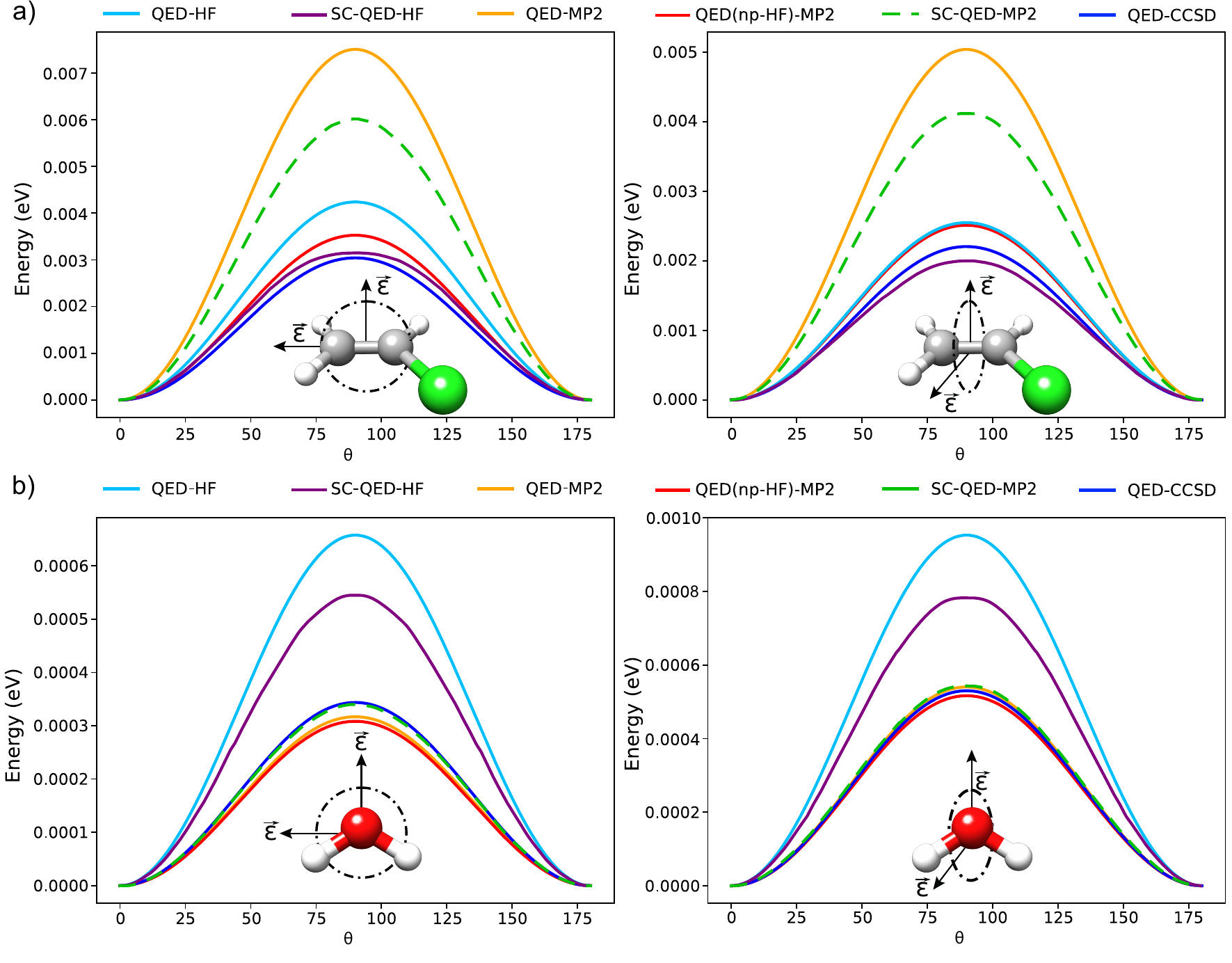}
    \caption{Field polarization $\pmb{\epsilon}$ orientational effects on chloroethylene (a) and water (b) inside an optical cavity with frequency and the light-matter coupling set to $\omega=$ \qtylist{2.72}{\electronvolt} and $\lambda=$ \qtylist{0.01}{\atomicunit} For both molecules, two orthogonal rotations of the field polarization are shown.}
    \label{orientational_effects}
\end{figure}
In Figure \ref{orientational_effects}, we investigate the orientational effects of the field polarization $\pmb{\epsilon}$ for chloroethylene (a) and water (b).
For both molecular systems we perform two orthogonal rotations of the polarization vector $\pmb{\epsilon}$ (left and right plots).
The cavity frequency and the light-matter coupling are set to $\omega=$ \qtylist{2.72}{\electronvolt} and $\lambda=$ \qtylist{0.01}{\atomicunit}
The offset of the curves is chosen such that they all start and end at zero energy.
In all the four cases we observe that the $\theta$-dispersions reach a maximum when the polarization lies in the molecular plane.
The closer the polarization vector is to the molecular plane, the orientational effects are described less accurately by the mean-field and perturbative approaches.
For chloroethylene (a) the SC-QED-HF performs the best, while all the other approaches overestimate the orientational effects. Among the perturbative methods, the QED(np-HF)-MP2 outperforms the others.
For water (b), instead, we observe that the perturbative approaches perform collectively better with respect to the mean-field methods.
In particular, when the rotation is parallel to the $\mathrm{C}_2$ symmetry axis, the SC-QED-MP2 performs best. In the orthogonal case, instead, the QED-MP2 and QED(np-HF)-MP2 are closer to the QED-CCSD curve.
These orientational effects are purely polaritonic and come from the interplay between DSE contributions and electron-photon correlation.
The DSE gives a positive contribution to the energy reaching its maximum when the cavity polarization and the largest polarizability principal axis are aligned. On the other hand, the electron-photon correlation has a negative contribution to the energy such that the overall observed behavior results from cancellation of these two effects.

\section{4. Conclusions}
In this study, building upon the work of Bauer \textit{et al.},~\cite{bauer2023perturbation} we have developed of a M{\o}ller-Plesset perturbation theory based on the reliable SC-QED-HF polaritonic molecular orbitals.~\cite{riso2022molecular,el2024toward}
Our analysis reveals that employing a fully consistent molecular orbital framework for the zeroth-order Hamiltonian is critical to effectively capture the cavity-induced electron-photon correlation effects.
For instance, the cavity coupling and frequency dispersions are well reproduced by the SC-QED-MP2 approach which includes some correlation effects at the mean-field level.
On the other hand, the QED-MP2 and QED(np-HF)-MP2 methods capture the electron-photon correlation only perturbatively, and higher orders in perturbation theory may be needed in order to obtain higher accuracy.
However, M{\o}ller-Plesset perturbation theory is not guaranteed to converge and further investigations of the converge properties in the polaritonic Hilbert space $\mathcal{H}_{pol} = \mathcal{H}_e \otimes \mathcal{H}_{ph}$ may provide interesting numerical insights.~\cite{olsen2000divergence,roden2024perturbative,forsberg2000convergence}
The use of a correct molecular orbital theory is crucial in order to avoid unphysical behavior, such as the ones displayed by QED-MP2 and LF-MP2 in the long-range regime of intermolecular interactions.
We also point out that the generalization to an \textit{ab initio} multi-mode QED Hamiltonian is less trivial than just having a diagonal transformation for each mode.
Care must be taken in order to obtain a multi-mode molecular orbital theory.
Efforts in this direction are currently under way.
This work paves the way for the development of more accurate perturbative approaches, such as QED versions of CC2 and CC3. Similarly, active space methodologies can only be extended to QED environments for fully consistent molecular orbital theories.
Continuous advancements in experimental setups to achieve larger coupling strengths may advocate for using SC-QED-MP2 in forthcoming studies.

\begin{acknowledgement}
Y.E.M., R.R.R., M.C. and H.K. acknowledge funding from the European Research Council (ERC) under the European Union’s Horizon 2020 Research and Innovation Programme (grant agreement No. 101020016).
E.R. acknowledges funding from the European Research Council (ERC) under the European Union’s Horizon Europe Research and Innovation Programme (Grant No. ERC-StG-2021-101040197—QED-Spin).
\end{acknowledgement}
\begin{suppinfo}
Development of general QED M{\o}ller-Plesset perturbation theory for multiple set of bosons and modes. Derivation of strong coupling QED M{\o}ller-Plesset perturbation theory.
\end{suppinfo}
\section{Code availability} 
The development version of the e$^\mathcal{T}$ program\cite{folkestad2020t} used to perform the calculations shown in this work is available from the corresponding author upon reasonable request.
\section{Data availability} 
The data of this work are available at \doi{10.5281/zenodo.14639372}.

\bibliography{Main}

\providecommand{\latin}[1]{#1}
\makeatletter
\providecommand{\doi}
  {\begingroup\let\do\@makeother\dospecials
  \catcode`\{=1 \catcode`\}=2 \doi@aux}
\providecommand{\doi@aux}[1]{\endgroup\texttt{#1}}
\makeatother
\providecommand*\mcitethebibliography{\thebibliography}
\csname @ifundefined\endcsname{endmcitethebibliography}
  {\let\endmcitethebibliography\endthebibliography}{}
\begin{mcitethebibliography}{68}
\providecommand*\natexlab[1]{#1}
\providecommand*\mciteSetBstSublistMode[1]{}
\providecommand*\mciteSetBstMaxWidthForm[2]{}
\providecommand*\mciteBstWouldAddEndPuncttrue
  {\def\EndOfBibitem{\unskip.}}
\providecommand*\mciteBstWouldAddEndPunctfalse
  {\let\EndOfBibitem\relax}
\providecommand*\mciteSetBstMidEndSepPunct[3]{}
\providecommand*\mciteSetBstSublistLabelBeginEnd[3]{}
\providecommand*\EndOfBibitem{}
\mciteSetBstSublistMode{f}
\mciteSetBstMaxWidthForm{subitem}{(\alph{mcitesubitemcount})}
\mciteSetBstSublistLabelBeginEnd
  {\mcitemaxwidthsubitemform\space}
  {\relax}
  {\relax}

\bibitem[Garcia-Vidal \latin{et~al.}(2021)Garcia-Vidal, Ciuti, and
  Ebbesen]{garcia2021manipulating}
Garcia-Vidal,~F.~J.; Ciuti,~C.; Ebbesen,~T.~W. Manipulating matter by strong
  coupling to vacuum fields. \emph{Science} \textbf{2021}, \emph{373},
  eabd0336\relax
\mciteBstWouldAddEndPuncttrue
\mciteSetBstMidEndSepPunct{\mcitedefaultmidpunct}
{\mcitedefaultendpunct}{\mcitedefaultseppunct}\relax
\EndOfBibitem
\bibitem[Flick \latin{et~al.}(2018)Flick, Rivera, and Narang]{flick2018strong}
Flick,~J.; Rivera,~N.; Narang,~P. Strong light-matter coupling in quantum
  chemistry and quantum photonics. \emph{Nanophotonics} \textbf{2018},
  \emph{7}, 1479--1501\relax
\mciteBstWouldAddEndPuncttrue
\mciteSetBstMidEndSepPunct{\mcitedefaultmidpunct}
{\mcitedefaultendpunct}{\mcitedefaultseppunct}\relax
\EndOfBibitem
\bibitem[Herrera and Owrutsky(2020)Herrera, and Owrutsky]{herrera2020molecular}
Herrera,~F.; Owrutsky,~J. Molecular polaritons for controlling chemistry with
  quantum optics. \emph{J. Chem. Phys.} \textbf{2020}, \emph{152}, 100902\relax
\mciteBstWouldAddEndPuncttrue
\mciteSetBstMidEndSepPunct{\mcitedefaultmidpunct}
{\mcitedefaultendpunct}{\mcitedefaultseppunct}\relax
\EndOfBibitem
\bibitem[Sandik \latin{et~al.}(2024)Sandik, Feist, Garc{\'\i}a-Vidal, and
  Schwartz]{sandik2024cavity}
Sandik,~G.; Feist,~J.; Garc{\'\i}a-Vidal,~F.~J.; Schwartz,~T. Cavity-enhanced
  energy transport in molecular systems. \emph{Nat. Mater.} \textbf{2024},
  1--12\relax
\mciteBstWouldAddEndPuncttrue
\mciteSetBstMidEndSepPunct{\mcitedefaultmidpunct}
{\mcitedefaultendpunct}{\mcitedefaultseppunct}\relax
\EndOfBibitem
\bibitem[Fukushima \latin{et~al.}(2022)Fukushima, Yoshimitsu, and
  Murakoshi]{fukushima2022inherent}
Fukushima,~T.; Yoshimitsu,~S.; Murakoshi,~K. Inherent Promotion of Ionic
  Conductivity via Collective Vibrational Strong Coupling of Water with the
  Vacuum Electromagnetic Field. \emph{J. Am. Chem. Soc.} \textbf{2022},
  \emph{144}, 12177--12183\relax
\mciteBstWouldAddEndPuncttrue
\mciteSetBstMidEndSepPunct{\mcitedefaultmidpunct}
{\mcitedefaultendpunct}{\mcitedefaultseppunct}\relax
\EndOfBibitem
\bibitem[Schachenmayer \latin{et~al.}(2015)Schachenmayer, Genes, Tignone, and
  Pupillo]{schachenmayer2015cavity}
Schachenmayer,~J.; Genes,~C.; Tignone,~E.; Pupillo,~G. Cavity-enhanced
  transport of excitons. \emph{Phys. Rev. Lett.} \textbf{2015}, \emph{114},
  196403\relax
\mciteBstWouldAddEndPuncttrue
\mciteSetBstMidEndSepPunct{\mcitedefaultmidpunct}
{\mcitedefaultendpunct}{\mcitedefaultseppunct}\relax
\EndOfBibitem
\bibitem[Mondal \latin{et~al.}(2022)Mondal, Semenov, Ochoa, and
  Nitzan]{mondal2022strong}
Mondal,~M.; Semenov,~A.; Ochoa,~M.~A.; Nitzan,~A. Strong Coupling in Infrared
  Plasmonic Cavities. \emph{J. Phys. Chem. Lett.} \textbf{2022}, \emph{13},
  9673--9678\relax
\mciteBstWouldAddEndPuncttrue
\mciteSetBstMidEndSepPunct{\mcitedefaultmidpunct}
{\mcitedefaultendpunct}{\mcitedefaultseppunct}\relax
\EndOfBibitem
\bibitem[Fitzgerald \latin{et~al.}(2016)Fitzgerald, Narang, Craster, Maier, and
  Giannini]{fitzgerald2016quantum}
Fitzgerald,~J.~M.; Narang,~P.; Craster,~R.~V.; Maier,~S.~A.; Giannini,~V.
  Quantum plasmonics. \emph{Proc. IEEE} \textbf{2016}, \emph{104},
  2307--2322\relax
\mciteBstWouldAddEndPuncttrue
\mciteSetBstMidEndSepPunct{\mcitedefaultmidpunct}
{\mcitedefaultendpunct}{\mcitedefaultseppunct}\relax
\EndOfBibitem
\bibitem[H{\"u}bener \latin{et~al.}(2021)H{\"u}bener, De~Giovannini,
  Sch{\"a}fer, Andberger, Ruggenthaler, Faist, and
  Rubio]{hubener2021engineering}
H{\"u}bener,~H.; De~Giovannini,~U.; Sch{\"a}fer,~C.; Andberger,~J.;
  Ruggenthaler,~M.; Faist,~J.; Rubio,~A. Engineering quantum materials with
  chiral optical cavities. \emph{Nat. Mater.} \textbf{2021}, \emph{20},
  438--442\relax
\mciteBstWouldAddEndPuncttrue
\mciteSetBstMidEndSepPunct{\mcitedefaultmidpunct}
{\mcitedefaultendpunct}{\mcitedefaultseppunct}\relax
\EndOfBibitem
\bibitem[Scholes(2020)]{scholes2020polaritons}
Scholes,~G.~D. Polaritons and excitons: Hamiltonian design for enhanced
  coherence. \emph{Proc. R. Soc. A} \textbf{2020}, \emph{476}, 20200278\relax
\mciteBstWouldAddEndPuncttrue
\mciteSetBstMidEndSepPunct{\mcitedefaultmidpunct}
{\mcitedefaultendpunct}{\mcitedefaultseppunct}\relax
\EndOfBibitem
\bibitem[Mauro \latin{et~al.}(2024)Mauro, Fregoni, Feist, and
  Avriller]{mauro2024classical}
Mauro,~L.; Fregoni,~J.; Feist,~J.; Avriller,~R. Classical approaches to chiral
  polaritonics. \emph{Phys. Rev. A} \textbf{2024}, \emph{109}, 023528\relax
\mciteBstWouldAddEndPuncttrue
\mciteSetBstMidEndSepPunct{\mcitedefaultmidpunct}
{\mcitedefaultendpunct}{\mcitedefaultseppunct}\relax
\EndOfBibitem
\bibitem[Sasaki \latin{et~al.}(2025)Sasaki, Takahashi, Taemaitree, Nakamura,
  Hutchison, Uji-i, and Hirai]{sasaki2025optical}
Sasaki,~I.; Takahashi,~K.; Taemaitree,~F.; Nakamura,~T.; Hutchison,~J.~A.;
  Uji-i,~H.; Hirai,~K. Optical Cavity Enhancement of Visible Light-Driven
  Photochemical Reaction in the Crystalline State. \emph{Chem. Commun.}
  \textbf{2025}, \relax
\mciteBstWouldAddEndPunctfalse
\mciteSetBstMidEndSepPunct{\mcitedefaultmidpunct}
{}{\mcitedefaultseppunct}\relax
\EndOfBibitem
\bibitem[Ribeiro \latin{et~al.}(2018)Ribeiro, Mart{\'\i}nez-Mart{\'\i}nez, Du,
  Campos-Gonzalez-Angulo, and Yuen-Zhou]{ribeiro2018polariton}
Ribeiro,~R.~F.; Mart{\'\i}nez-Mart{\'\i}nez,~L.~A.; Du,~M.;
  Campos-Gonzalez-Angulo,~J.; Yuen-Zhou,~J. Polariton chemistry: controlling
  molecular dynamics with optical cavities. \emph{Chem. Sci.} \textbf{2018},
  \emph{9}, 6325--6339\relax
\mciteBstWouldAddEndPuncttrue
\mciteSetBstMidEndSepPunct{\mcitedefaultmidpunct}
{\mcitedefaultendpunct}{\mcitedefaultseppunct}\relax
\EndOfBibitem
\bibitem[Fregoni \latin{et~al.}(2020)Fregoni, Granucci, Persico, and
  Corni]{fregoni2020strong}
Fregoni,~J.; Granucci,~G.; Persico,~M.; Corni,~S. Strong coupling with light
  enhances the photoisomerization quantum yield of azobenzene. \emph{Chem}
  \textbf{2020}, \emph{6}, 250--265\relax
\mciteBstWouldAddEndPuncttrue
\mciteSetBstMidEndSepPunct{\mcitedefaultmidpunct}
{\mcitedefaultendpunct}{\mcitedefaultseppunct}\relax
\EndOfBibitem
\bibitem[Lather \latin{et~al.}(2019)Lather, Bhatt, Thomas, Ebbesen, and
  George]{lather2019cavity}
Lather,~J.; Bhatt,~P.; Thomas,~A.; Ebbesen,~T.~W.; George,~J. Cavity catalysis
  by cooperative vibrational strong coupling of reactant and solvent molecules.
  \emph{Angew.Chem.Int.Ed.} \textbf{2019}, \emph{58}, 10635--10638\relax
\mciteBstWouldAddEndPuncttrue
\mciteSetBstMidEndSepPunct{\mcitedefaultmidpunct}
{\mcitedefaultendpunct}{\mcitedefaultseppunct}\relax
\EndOfBibitem
\bibitem[Ahn \latin{et~al.}(2023)Ahn, Triana, Recabal, Herrera, and
  Simpkins]{ahn2023modification}
Ahn,~W.; Triana,~J.~F.; Recabal,~F.; Herrera,~F.; Simpkins,~B.~S. Modification
  of ground-state chemical reactivity via light--matter coherence in infrared
  cavities. \emph{Science} \textbf{2023}, \emph{380}, 1165--1168\relax
\mciteBstWouldAddEndPuncttrue
\mciteSetBstMidEndSepPunct{\mcitedefaultmidpunct}
{\mcitedefaultendpunct}{\mcitedefaultseppunct}\relax
\EndOfBibitem
\bibitem[Mony \latin{et~al.}(2021)Mony, Climent, Petersen, Moth-Poulsen, Feist,
  and B{\"o}rjesson]{mony2021photoisomerization}
Mony,~J.; Climent,~C.; Petersen,~A.~U.; Moth-Poulsen,~K.; Feist,~J.;
  B{\"o}rjesson,~K. Photoisomerization efficiency of a solar thermal fuel in
  the strong coupling regime. \emph{Adv. Funct. Mater.} \textbf{2021},
  \emph{31}, 2010737\relax
\mciteBstWouldAddEndPuncttrue
\mciteSetBstMidEndSepPunct{\mcitedefaultmidpunct}
{\mcitedefaultendpunct}{\mcitedefaultseppunct}\relax
\EndOfBibitem
\bibitem[Mandal \latin{et~al.}(2023)Mandal, Taylor, and Huo]{mandal2023theory}
Mandal,~A.; Taylor,~M.~A.; Huo,~P. Theory for Cavity-Modified Ground-State
  Reactivities via Electron--Photon Interactions. \emph{J. Phys. Chem. A}
  \textbf{2023}, \emph{127}, 6830--6841\relax
\mciteBstWouldAddEndPuncttrue
\mciteSetBstMidEndSepPunct{\mcitedefaultmidpunct}
{\mcitedefaultendpunct}{\mcitedefaultseppunct}\relax
\EndOfBibitem
\bibitem[Fischer \latin{et~al.}(2022)Fischer, Anders, and
  Saalfrank]{fischer2022cavity}
Fischer,~E.~W.; Anders,~J.; Saalfrank,~P. Cavity-altered thermal isomerization
  rates and dynamical resonant localization in vibro-polaritonic chemistry.
  \emph{J. Chem. Phys.} \textbf{2022}, \emph{156}, 154305\relax
\mciteBstWouldAddEndPuncttrue
\mciteSetBstMidEndSepPunct{\mcitedefaultmidpunct}
{\mcitedefaultendpunct}{\mcitedefaultseppunct}\relax
\EndOfBibitem
\bibitem[Krupp \latin{et~al.}(2024)Krupp, Groenhof, and
  Vendrell]{krupp2024quantum}
Krupp,~N.; Groenhof,~G.; Vendrell,~O. Quantum dynamics simulation of
  exciton-polariton transport. \emph{arXiv preprint arXiv:2410.23739}
  \textbf{2024}, \relax
\mciteBstWouldAddEndPunctfalse
\mciteSetBstMidEndSepPunct{\mcitedefaultmidpunct}
{}{\mcitedefaultseppunct}\relax
\EndOfBibitem
\bibitem[Chikkaraddy \latin{et~al.}(2016)Chikkaraddy, De~Nijs, Benz, Barrow,
  Scherman, Rosta, Demetriadou, Fox, Hess, and Baumberg]{chikkaraddy2016single}
Chikkaraddy,~R.; De~Nijs,~B.; Benz,~F.; Barrow,~S.~J.; Scherman,~O.~A.;
  Rosta,~E.; Demetriadou,~A.; Fox,~P.; Hess,~O.; Baumberg,~J.~J.
  Single-molecule strong coupling at room temperature in plasmonic
  nanocavities. \emph{Nature} \textbf{2016}, \emph{535}, 127--130\relax
\mciteBstWouldAddEndPuncttrue
\mciteSetBstMidEndSepPunct{\mcitedefaultmidpunct}
{\mcitedefaultendpunct}{\mcitedefaultseppunct}\relax
\EndOfBibitem
\bibitem[Santhosh \latin{et~al.}(2016)Santhosh, Bitton, Chuntonov, and
  Haran]{santhosh2016vacuum}
Santhosh,~K.; Bitton,~O.; Chuntonov,~L.; Haran,~G. Vacuum Rabi splitting in a
  plasmonic cavity at the single quantum emitter limit. \emph{Nat. Commun.}
  \textbf{2016}, \emph{7}, ncomms11823\relax
\mciteBstWouldAddEndPuncttrue
\mciteSetBstMidEndSepPunct{\mcitedefaultmidpunct}
{\mcitedefaultendpunct}{\mcitedefaultseppunct}\relax
\EndOfBibitem
\bibitem[Bitton and Haran(2022)Bitton, and Haran]{bitton2022plasmonic}
Bitton,~O.; Haran,~G. Plasmonic Cavities and Individual Quantum Emitters in the
  Strong Coupling Limit. \emph{Acc. Chem. Res.} \textbf{2022}, \emph{55},
  1659--1668\relax
\mciteBstWouldAddEndPuncttrue
\mciteSetBstMidEndSepPunct{\mcitedefaultmidpunct}
{\mcitedefaultendpunct}{\mcitedefaultseppunct}\relax
\EndOfBibitem
\bibitem[Baumberg(2022)]{baumberg2022picocavities}
Baumberg,~J.~J. Picocavities: A primer. \emph{Nano Lett.} \textbf{2022},
  \emph{22}, 5859--5865\relax
\mciteBstWouldAddEndPuncttrue
\mciteSetBstMidEndSepPunct{\mcitedefaultmidpunct}
{\mcitedefaultendpunct}{\mcitedefaultseppunct}\relax
\EndOfBibitem
\bibitem[Damari \latin{et~al.}(2019)Damari, Weinberg, Krotkov, Demina, Akulov,
  Golombek, Schwartz, and Fleischer]{damari2019strong}
Damari,~R.; Weinberg,~O.; Krotkov,~D.; Demina,~N.; Akulov,~K.; Golombek,~A.;
  Schwartz,~T.; Fleischer,~S. Strong coupling of collective intermolecular
  vibrations in organic materials at terahertz frequencies. \emph{Nat. Commun.}
  \textbf{2019}, \emph{10}, 3248\relax
\mciteBstWouldAddEndPuncttrue
\mciteSetBstMidEndSepPunct{\mcitedefaultmidpunct}
{\mcitedefaultendpunct}{\mcitedefaultseppunct}\relax
\EndOfBibitem
\bibitem[Castagnola \latin{et~al.}(2024)Castagnola, Haugland, Ronca, Koch, and
  Sch\"{a}fer]{castagnola2024collective}
Castagnola,~M.; Haugland,~T.~S.; Ronca,~E.; Koch,~H.; Sch\"{a}fer,~C.
  Collective strong coupling modifies aggregation and solvation. \emph{J. Phys.
  Chem. Lett.} \textbf{2024}, \emph{15}, 1428--1434\relax
\mciteBstWouldAddEndPuncttrue
\mciteSetBstMidEndSepPunct{\mcitedefaultmidpunct}
{\mcitedefaultendpunct}{\mcitedefaultseppunct}\relax
\EndOfBibitem
\bibitem[Ashida \latin{et~al.}(2021)Ashida, {\.I}mamo{\u{g}}lu, and
  Demler]{ashida2021cavity}
Ashida,~Y.; {\.I}mamo{\u{g}}lu,~A.; Demler,~E. Cavity quantum electrodynamics
  at arbitrary light-matter coupling strengths. \emph{Phys. Rev. Lett.}
  \textbf{2021}, \emph{126}, 153603\relax
\mciteBstWouldAddEndPuncttrue
\mciteSetBstMidEndSepPunct{\mcitedefaultmidpunct}
{\mcitedefaultendpunct}{\mcitedefaultseppunct}\relax
\EndOfBibitem
\bibitem[Haugland \latin{et~al.}(2020)Haugland, Ronca, Kj{\o}nstad, Rubio, and
  Koch]{haugland2020coupled}
Haugland,~T.~S.; Ronca,~E.; Kj{\o}nstad,~E.~F.; Rubio,~A.; Koch,~H. Coupled
  cluster theory for molecular polaritons: Changing ground and excited states.
  \emph{Phys.l Rev. X} \textbf{2020}, \emph{10}, 041043\relax
\mciteBstWouldAddEndPuncttrue
\mciteSetBstMidEndSepPunct{\mcitedefaultmidpunct}
{\mcitedefaultendpunct}{\mcitedefaultseppunct}\relax
\EndOfBibitem
\bibitem[Taylor \latin{et~al.}(2020)Taylor, Mandal, Zhou, and
  Huo]{taylor2020resolution}
Taylor,~M.~A.; Mandal,~A.; Zhou,~W.; Huo,~P. Resolution of gauge ambiguities in
  molecular cavity quantum electrodynamics. \emph{Phys. Rev. Lett.}
  \textbf{2020}, \emph{125}, 123602\relax
\mciteBstWouldAddEndPuncttrue
\mciteSetBstMidEndSepPunct{\mcitedefaultmidpunct}
{\mcitedefaultendpunct}{\mcitedefaultseppunct}\relax
\EndOfBibitem
\bibitem[McTague and Foley(2022)McTague, and Foley]{mctague2022non}
McTague,~J.; Foley,~J.~J. Non-Hermitian cavity quantum
  electrodynamics--configuration interaction singles approach for polaritonic
  structure with ab initio molecular Hamiltonians. \emph{J. Chem. Phys.}
  \textbf{2022}, \emph{156}\relax
\mciteBstWouldAddEndPuncttrue
\mciteSetBstMidEndSepPunct{\mcitedefaultmidpunct}
{\mcitedefaultendpunct}{\mcitedefaultseppunct}\relax
\EndOfBibitem
\bibitem[Du and Yuen-Zhou(2022)Du, and Yuen-Zhou]{du2022catalysis}
Du,~M.; Yuen-Zhou,~J. Catalysis by dark states in vibropolaritonic chemistry.
  \emph{Phys. Rev. Lett.} \textbf{2022}, \emph{128}, 096001\relax
\mciteBstWouldAddEndPuncttrue
\mciteSetBstMidEndSepPunct{\mcitedefaultmidpunct}
{\mcitedefaultendpunct}{\mcitedefaultseppunct}\relax
\EndOfBibitem
\bibitem[Huang and Liang(2025)Huang, and Liang]{huang2025analytical}
Huang,~X.; Liang,~W. Analytical derivative approaches for vibro-polaritonic
  structures and properties. I. Formalism and implementation. \emph{J. Chem.
  Phys.} \textbf{2025}, \emph{162}\relax
\mciteBstWouldAddEndPuncttrue
\mciteSetBstMidEndSepPunct{\mcitedefaultmidpunct}
{\mcitedefaultendpunct}{\mcitedefaultseppunct}\relax
\EndOfBibitem
\bibitem[Ruggenthaler \latin{et~al.}(2014)Ruggenthaler, Flick, Pellegrini,
  Appel, Tokatly, and Rubio]{ruggenthaler2014quantum}
Ruggenthaler,~M.; Flick,~J.; Pellegrini,~C.; Appel,~H.; Tokatly,~I.~V.;
  Rubio,~A. Quantum-electrodynamical density-functional theory: Bridging
  quantum optics and electronic-structure theory. \emph{Phys. Rev. A}
  \textbf{2014}, \emph{90}, 012508\relax
\mciteBstWouldAddEndPuncttrue
\mciteSetBstMidEndSepPunct{\mcitedefaultmidpunct}
{\mcitedefaultendpunct}{\mcitedefaultseppunct}\relax
\EndOfBibitem
\bibitem[Sch{\"a}fer \latin{et~al.}(2022)Sch{\"a}fer, Flick, Ronca, Narang, and
  Rubio]{schafer2022shining}
Sch{\"a}fer,~C.; Flick,~J.; Ronca,~E.; Narang,~P.; Rubio,~A. Shining light on
  the microscopic resonant mechanism responsible for cavity-mediated chemical
  reactivity. \emph{Nat. Commun.} \textbf{2022}, \emph{13}, 7817\relax
\mciteBstWouldAddEndPuncttrue
\mciteSetBstMidEndSepPunct{\mcitedefaultmidpunct}
{\mcitedefaultendpunct}{\mcitedefaultseppunct}\relax
\EndOfBibitem
\bibitem[Sch{\"a}fer \latin{et~al.}(2018)Sch{\"a}fer, Ruggenthaler, and
  Rubio]{schafer2018ab}
Sch{\"a}fer,~C.; Ruggenthaler,~M.; Rubio,~A. Ab initio nonrelativistic quantum
  electrodynamics: Bridging quantum chemistry and quantum optics from weak to
  strong coupling. \emph{Phys. Rev. A} \textbf{2018}, \emph{98}, 043801\relax
\mciteBstWouldAddEndPuncttrue
\mciteSetBstMidEndSepPunct{\mcitedefaultmidpunct}
{\mcitedefaultendpunct}{\mcitedefaultseppunct}\relax
\EndOfBibitem
\bibitem[Flick \latin{et~al.}(2018)Flick, Sc{\"a}hfer, Ruggenthaler, Appel, and
  Rubio]{flick2018ab}
Flick,~J.; Sc{\"a}hfer,~C.; Ruggenthaler,~M.; Appel,~H.; Rubio,~A. Ab initio
  optimized effective potentials for real molecules in optical cavities: Photon
  contributions to the molecular ground state. \emph{ACS photonics}
  \textbf{2018}, \emph{5}, 992--1005\relax
\mciteBstWouldAddEndPuncttrue
\mciteSetBstMidEndSepPunct{\mcitedefaultmidpunct}
{\mcitedefaultendpunct}{\mcitedefaultseppunct}\relax
\EndOfBibitem
\bibitem[Riso \latin{et~al.}(2022)Riso, Haugland, Ronca, and
  Koch]{riso2022characteristic}
Riso,~R.~R.; Haugland,~T.~S.; Ronca,~E.; Koch,~H. On the characteristic
  features of ionization in QED environments. \emph{J. Chem. Phys.}
  \textbf{2022}, \emph{156}, 234103\relax
\mciteBstWouldAddEndPuncttrue
\mciteSetBstMidEndSepPunct{\mcitedefaultmidpunct}
{\mcitedefaultendpunct}{\mcitedefaultseppunct}\relax
\EndOfBibitem
\bibitem[Pavosevic \latin{et~al.}(2022)Pavosevic, Hammes-Schiffer, Rubio, and
  Flick]{pavosevic2022cavity}
Pavosevic,~F.; Hammes-Schiffer,~S.; Rubio,~A.; Flick,~J. Cavity-modulated
  proton transfer reactions. \emph{J. Am. Chem. Soc.} \textbf{2022},
  \emph{144}, 4995--5002\relax
\mciteBstWouldAddEndPuncttrue
\mciteSetBstMidEndSepPunct{\mcitedefaultmidpunct}
{\mcitedefaultendpunct}{\mcitedefaultseppunct}\relax
\EndOfBibitem
\bibitem[DePrince(2021)]{deprince2021cavity}
DePrince,~A.~E. Cavity-modulated ionization potentials and electron affinities
  from quantum electrodynamics coupled-cluster theory. \emph{J. Chem. Phys.}
  \textbf{2021}, \emph{154}\relax
\mciteBstWouldAddEndPuncttrue
\mciteSetBstMidEndSepPunct{\mcitedefaultmidpunct}
{\mcitedefaultendpunct}{\mcitedefaultseppunct}\relax
\EndOfBibitem
\bibitem[Liebenthal \latin{et~al.}(2022)Liebenthal, Vu, and
  DePrince]{liebenthal2022equation}
Liebenthal,~M.~D.; Vu,~N.; DePrince,~A.~E. Equation-of-motion cavity quantum
  electrodynamics coupled-cluster theory for electron attachment. \emph{J.
  Chem. Phys.} \textbf{2022}, \emph{156}\relax
\mciteBstWouldAddEndPuncttrue
\mciteSetBstMidEndSepPunct{\mcitedefaultmidpunct}
{\mcitedefaultendpunct}{\mcitedefaultseppunct}\relax
\EndOfBibitem
\bibitem[Mordovina \latin{et~al.}(2020)Mordovina, Bungey, Appel, Knowles,
  Rubio, and Manby]{mordovina2020polaritonic}
Mordovina,~U.; Bungey,~C.; Appel,~H.; Knowles,~P.~J.; Rubio,~A.; Manby,~F.~R.
  Polaritonic coupled-cluster theory. \emph{Phys. Rev. Res.} \textbf{2020},
  \emph{2}, 023262\relax
\mciteBstWouldAddEndPuncttrue
\mciteSetBstMidEndSepPunct{\mcitedefaultmidpunct}
{\mcitedefaultendpunct}{\mcitedefaultseppunct}\relax
\EndOfBibitem
\bibitem[Helgaker \latin{et~al.}(2013)Helgaker, Jorgensen, and
  Olsen]{helgaker2013molecular}
Helgaker,~T.; Jorgensen,~P.; Olsen,~J. \emph{Molecular electronic-structure
  theory}; John Wiley \& Sons, 2013\relax
\mciteBstWouldAddEndPuncttrue
\mciteSetBstMidEndSepPunct{\mcitedefaultmidpunct}
{\mcitedefaultendpunct}{\mcitedefaultseppunct}\relax
\EndOfBibitem
\bibitem[Haugland \latin{et~al.}(2023)Haugland, Philbin, Ghosh, Chen, Koch, and
  Narang]{haugland2023perturbation}
Haugland,~T.~S.; Philbin,~J.~P.; Ghosh,~T.~K.; Chen,~M.; Koch,~H.; Narang,~P.
  Understanding the polaritonic ground state in cavity quantum electrodynamics.
  \emph{arXiv preprint arXiv:2307.14822} \textbf{2023}, \relax
\mciteBstWouldAddEndPunctfalse
\mciteSetBstMidEndSepPunct{\mcitedefaultmidpunct}
{}{\mcitedefaultseppunct}\relax
\EndOfBibitem
\bibitem[Bauer and Dreuw(2023)Bauer, and Dreuw]{bauer2023perturbation}
Bauer,~M.; Dreuw,~A. Perturbation theoretical approaches to strong
  light--matter coupling in ground and excited electronic states for the
  description of molecular polaritons. \emph{J. Chem. Phys.} \textbf{2023},
  \emph{158}\relax
\mciteBstWouldAddEndPuncttrue
\mciteSetBstMidEndSepPunct{\mcitedefaultmidpunct}
{\mcitedefaultendpunct}{\mcitedefaultseppunct}\relax
\EndOfBibitem
\bibitem[Riso \latin{et~al.}(2022)Riso, Haugland, Ronca, and
  Koch]{riso2022molecular}
Riso,~R.~R.; Haugland,~T.~S.; Ronca,~E.; Koch,~H. Molecular orbital theory in
  cavity QED environments. \emph{Nat. Commun.} \textbf{2022}, \emph{13},
  1368\relax
\mciteBstWouldAddEndPuncttrue
\mciteSetBstMidEndSepPunct{\mcitedefaultmidpunct}
{\mcitedefaultendpunct}{\mcitedefaultseppunct}\relax
\EndOfBibitem
\bibitem[El~Moutaoukal \latin{et~al.}(2024)El~Moutaoukal, Riso, Castagnola, and
  Koch]{el2024toward}
El~Moutaoukal,~Y.; Riso,~R.~R.; Castagnola,~M.; Koch,~H. Toward polaritonic
  molecular orbitals for large molecular systems. \emph{J. Chem. Theory
  Comput.} \textbf{2024}, \emph{20}, 8911--8920\relax
\mciteBstWouldAddEndPuncttrue
\mciteSetBstMidEndSepPunct{\mcitedefaultmidpunct}
{\mcitedefaultendpunct}{\mcitedefaultseppunct}\relax
\EndOfBibitem
\bibitem[Cui \latin{et~al.}(2024)Cui, Mandal, and Reichman]{cui2024variational}
Cui,~Z.-H.; Mandal,~A.; Reichman,~D.~R. Variational lang--firsov approach plus
  m{\o}ller--plesset perturbation theory with applications to ab initio
  polariton chemistry. \emph{J. Chem. Theory Comput.} \textbf{2024}, \emph{20},
  1143--1156\relax
\mciteBstWouldAddEndPuncttrue
\mciteSetBstMidEndSepPunct{\mcitedefaultmidpunct}
{\mcitedefaultendpunct}{\mcitedefaultseppunct}\relax
\EndOfBibitem
\bibitem[Cohen-Tannoudji \latin{et~al.}(2024)Cohen-Tannoudji, Dupont-Roc, and
  Grynberg]{cohen2024photons}
Cohen-Tannoudji,~C.; Dupont-Roc,~J.; Grynberg,~G. \emph{Photons and atoms:
  introduction to quantum electrodynamics}; John Wiley \& Sons, 2024\relax
\mciteBstWouldAddEndPuncttrue
\mciteSetBstMidEndSepPunct{\mcitedefaultmidpunct}
{\mcitedefaultendpunct}{\mcitedefaultseppunct}\relax
\EndOfBibitem
\bibitem[Mandal \latin{et~al.}(2020)Mandal, Montillo~Vega, and
  Huo]{mandal2020polarized}
Mandal,~A.; Montillo~Vega,~S.; Huo,~P. Polarized Fock states and the dynamical
  Casimir effect in molecular cavity quantum electrodynamics. \emph{J. Phys.
  Chem. Lett.} \textbf{2020}, \emph{11}, 9215--9223\relax
\mciteBstWouldAddEndPuncttrue
\mciteSetBstMidEndSepPunct{\mcitedefaultmidpunct}
{\mcitedefaultendpunct}{\mcitedefaultseppunct}\relax
\EndOfBibitem
\bibitem[De~Liberato(2014)]{de2014light}
De~Liberato,~S. Light-matter decoupling in the deep strong coupling regime: The
  breakdown of the Purcell effect. \emph{Phys. Rev. Lett.} \textbf{2014},
  \emph{112}, 016401\relax
\mciteBstWouldAddEndPuncttrue
\mciteSetBstMidEndSepPunct{\mcitedefaultmidpunct}
{\mcitedefaultendpunct}{\mcitedefaultseppunct}\relax
\EndOfBibitem
\bibitem[Di~Stefano \latin{et~al.}(2019)Di~Stefano, Settineri, Macr{\`\i},
  Garziano, Stassi, Savasta, and Nori]{di2019resolution}
Di~Stefano,~O.; Settineri,~A.; Macr{\`\i},~V.; Garziano,~L.; Stassi,~R.;
  Savasta,~S.; Nori,~F. Resolution of gauge ambiguities in ultrastrong-coupling
  cavity quantum electrodynamics. \emph{Nat. Phys.} \textbf{2019}, \emph{15},
  803--808\relax
\mciteBstWouldAddEndPuncttrue
\mciteSetBstMidEndSepPunct{\mcitedefaultmidpunct}
{\mcitedefaultendpunct}{\mcitedefaultseppunct}\relax
\EndOfBibitem
\bibitem[Frisk~Kockum \latin{et~al.}(2019)Frisk~Kockum, Miranowicz,
  De~Liberato, Savasta, and Nori]{frisk2019ultrastrong}
Frisk~Kockum,~A.; Miranowicz,~A.; De~Liberato,~S.; Savasta,~S.; Nori,~F.
  Ultrastrong coupling between light and matter. \emph{Nat. Rev. Phys.}
  \textbf{2019}, \emph{1}, 19--40\relax
\mciteBstWouldAddEndPuncttrue
\mciteSetBstMidEndSepPunct{\mcitedefaultmidpunct}
{\mcitedefaultendpunct}{\mcitedefaultseppunct}\relax
\EndOfBibitem
\bibitem[Rokaj \latin{et~al.}(2018)Rokaj, Welakuh, Ruggenthaler, and
  Rubio]{rokaj2018light}
Rokaj,~V.; Welakuh,~D.~M.; Ruggenthaler,~M.; Rubio,~A. Light--matter
  interaction in the long-wavelength limit: no ground-state without dipole
  self-energy. \emph{J. Phys. B: At. Mol. Opt. Phys.} \textbf{2018}, \emph{51},
  034005\relax
\mciteBstWouldAddEndPuncttrue
\mciteSetBstMidEndSepPunct{\mcitedefaultmidpunct}
{\mcitedefaultendpunct}{\mcitedefaultseppunct}\relax
\EndOfBibitem
\bibitem[L{\"o}wdin(1951)]{lowdin1951note}
L{\"o}wdin,~P.-O. A note on the quantum-mechanical perturbation theory.
  \emph{J. Chem. Phys.} \textbf{1951}, \emph{19}, 1396--1401\relax
\mciteBstWouldAddEndPuncttrue
\mciteSetBstMidEndSepPunct{\mcitedefaultmidpunct}
{\mcitedefaultendpunct}{\mcitedefaultseppunct}\relax
\EndOfBibitem
\bibitem[M{\o}ller and Plesset(1934)M{\o}ller, and Plesset]{moller1934note}
M{\o}ller,~C.; Plesset,~M.~S. Note on an approximation treatment for
  many-electron systems. \emph{Phys. Rev.} \textbf{1934}, \emph{46}, 618\relax
\mciteBstWouldAddEndPuncttrue
\mciteSetBstMidEndSepPunct{\mcitedefaultmidpunct}
{\mcitedefaultendpunct}{\mcitedefaultseppunct}\relax
\EndOfBibitem
\bibitem[Haugland \latin{et~al.}(2021)Haugland, Sch{\"a}fer, Ronca, Rubio, and
  Koch]{haugland2021intermolecular}
Haugland,~T.~S.; Sch{\"a}fer,~C.; Ronca,~E.; Rubio,~A.; Koch,~H. Intermolecular
  interactions in optical cavities: An ab initio QED study. \emph{J. Chem.
  Phys.} \textbf{2021}, \emph{154}, 094113\relax
\mciteBstWouldAddEndPuncttrue
\mciteSetBstMidEndSepPunct{\mcitedefaultmidpunct}
{\mcitedefaultendpunct}{\mcitedefaultseppunct}\relax
\EndOfBibitem
\bibitem[Craig and Thirunamachandran(1998)Craig, and
  Thirunamachandran]{craig1998molecular}
Craig,~D.~P.; Thirunamachandran,~T. \emph{Molecular quantum electrodynamics: an
  introduction to radiation-molecule interactions}; Courier Corporation,
  1998\relax
\mciteBstWouldAddEndPuncttrue
\mciteSetBstMidEndSepPunct{\mcitedefaultmidpunct}
{\mcitedefaultendpunct}{\mcitedefaultseppunct}\relax
\EndOfBibitem
\bibitem[Folkestad \latin{et~al.}(2020)Folkestad, Kj{\o}nstad, Myhre, Andersen,
  Balbi, Coriani, Giovannini, Goletto, Haugland, Hutcheson, \latin{et~al.}
  others]{folkestad2020t}
Folkestad,~S.~D.; Kj{\o}nstad,~E.~F.; Myhre,~R.~H.; Andersen,~J.~H.; Balbi,~A.;
  Coriani,~S.; Giovannini,~T.; Goletto,~L.; Haugland,~T.~S.; Hutcheson,~A.,
  \latin{et~al.}  e T 1.0: An open source electronic structure program with
  emphasis on coupled cluster and multilevel methods. \emph{J. Chem. Phys.}
  \textbf{2020}, \emph{152}, 184103\relax
\mciteBstWouldAddEndPuncttrue
\mciteSetBstMidEndSepPunct{\mcitedefaultmidpunct}
{\mcitedefaultendpunct}{\mcitedefaultseppunct}\relax
\EndOfBibitem
\bibitem[Cui(n.d.)]{cui_polar}
Cui,~Z.-H. {POLAR: Polariton and polaron (electron-boson coupled) systems from
  a quantum chemical perspective}.
  \url{https://github.com/zhcui/polar_preview}, n.d.; Accessed:
  2025-01-09\relax
\mciteBstWouldAddEndPuncttrue
\mciteSetBstMidEndSepPunct{\mcitedefaultmidpunct}
{\mcitedefaultendpunct}{\mcitedefaultseppunct}\relax
\EndOfBibitem
\bibitem[Sun \latin{et~al.}(2020)Sun, Zhang, Banerjee, Bao, Barbry, Blunt,
  Bogdanov, Booth, Chen, Cui, \latin{et~al.} others]{sun2020recent}
Sun,~Q.; Zhang,~X.; Banerjee,~S.; Bao,~P.; Barbry,~M.; Blunt,~N.~S.;
  Bogdanov,~N.~A.; Booth,~G.~H.; Chen,~J.; Cui,~Z.-H., \latin{et~al.}  Recent
  developments in the PySCF program package. \emph{J. Chem. Phys.}
  \textbf{2020}, \emph{153}\relax
\mciteBstWouldAddEndPuncttrue
\mciteSetBstMidEndSepPunct{\mcitedefaultmidpunct}
{\mcitedefaultendpunct}{\mcitedefaultseppunct}\relax
\EndOfBibitem
\bibitem[Pritchard \latin{et~al.}(2019)Pritchard, Altarawy, Didier, Gibsom, and
  Windus]{pritchard2019a}
Pritchard,~B.~P.; Altarawy,~D.; Didier,~B.; Gibsom,~T.~D.; Windus,~T.~L. A New
  Basis Set Exchange: An Open, Up-to-date Resource for the Molecular Sciences
  Community. \emph{J. Chem. Inf. Model.} \textbf{2019}, \emph{59},
  4814--4820\relax
\mciteBstWouldAddEndPuncttrue
\mciteSetBstMidEndSepPunct{\mcitedefaultmidpunct}
{\mcitedefaultendpunct}{\mcitedefaultseppunct}\relax
\EndOfBibitem
\bibitem[Dunning(1989)]{dunning1989a}
Dunning,~T.~H. Gaussian basis sets for use in correlated molecular
  calculations. I. The atoms boron through neon and hydrogen. \emph{J. Chem.
  Phys.} \textbf{1989}, \emph{90}, 1007--1023\relax
\mciteBstWouldAddEndPuncttrue
\mciteSetBstMidEndSepPunct{\mcitedefaultmidpunct}
{\mcitedefaultendpunct}{\mcitedefaultseppunct}\relax
\EndOfBibitem
\bibitem[Neese(2022)]{neese2022software}
Neese,~F. Software update: The ORCA program system—Version 5.0. \emph{Wiley
  Interdiscip. Rev.: Comput. Mol. Sci.} \textbf{2022}, \emph{12}, e1606\relax
\mciteBstWouldAddEndPuncttrue
\mciteSetBstMidEndSepPunct{\mcitedefaultmidpunct}
{\mcitedefaultendpunct}{\mcitedefaultseppunct}\relax
\EndOfBibitem
\bibitem[Ronca \latin{et~al.}(2014)Ronca, Belpassi, and
  Tarantelli]{ronca2014quantitative}
Ronca,~E.; Belpassi,~L.; Tarantelli,~F. A quantitative view of charge transfer
  in the hydrogen bond: the water dimer case. \emph{ChemPhysChem}
  \textbf{2014}, \emph{15}, 2682--2687\relax
\mciteBstWouldAddEndPuncttrue
\mciteSetBstMidEndSepPunct{\mcitedefaultmidpunct}
{\mcitedefaultendpunct}{\mcitedefaultseppunct}\relax
\EndOfBibitem
\bibitem[Olsen \latin{et~al.}(2000)Olsen, J{\o}rgensen, Helgaker, and
  Christiansen]{olsen2000divergence}
Olsen,~J.; J{\o}rgensen,~P.; Helgaker,~T.; Christiansen,~O. Divergence in
  M{\o}ller--Plesset theory: A simple explanation based on a two-state model.
  \emph{J. Chem. Phys.} \textbf{2000}, \emph{112}, 9736--9748\relax
\mciteBstWouldAddEndPuncttrue
\mciteSetBstMidEndSepPunct{\mcitedefaultmidpunct}
{\mcitedefaultendpunct}{\mcitedefaultseppunct}\relax
\EndOfBibitem
\bibitem[Roden and Foley(2024)Roden, and Foley]{roden2024perturbative}
Roden,~P.; Foley,~J.~J. Perturbative analysis of the coherent state
  transformation in ab initio cavity quantum electrodynamics. \emph{J. Chem.
  Phys.} \textbf{2024}, \emph{161}\relax
\mciteBstWouldAddEndPuncttrue
\mciteSetBstMidEndSepPunct{\mcitedefaultmidpunct}
{\mcitedefaultendpunct}{\mcitedefaultseppunct}\relax
\EndOfBibitem
\bibitem[Forsberg \latin{et~al.}(2000)Forsberg, He, He, and
  Cremer]{forsberg2000convergence}
Forsberg,~B.; He,~Z.; He,~Y.; Cremer,~D. Convergence behavior of the
  M{\o}ller--Plesset perturbation series: use of Feenberg scaling for the
  exclusion of backdoor intruder states. \emph{Int. J. Quantum Chem.}
  \textbf{2000}, \emph{76}, 306--330\relax
\mciteBstWouldAddEndPuncttrue
\mciteSetBstMidEndSepPunct{\mcitedefaultmidpunct}
{\mcitedefaultendpunct}{\mcitedefaultseppunct}\relax
\EndOfBibitem
\end{mcitethebibliography}




\section*{Supplementary material}

# Supporting information to:

# Strong coupling Møller-Plesset perturbation theory

Yassir El Moutaoukal,[†,‡] Rosario R. Riso,[†,‡] Matteo Castagnola,[†] Enrico Ronca,[¶] and Henrik Koch[*,†]

[†]Department of Chemistry, Norwegian University of Science and Technology, 7491 Trondheim, Norway

[‡]These authors contributed equally to this work

[¶]Department of Chemistry, Biology and Biotechnology, University of Perugia, Via Elce di Sotto, 8, 06123, Perugia, Italy

E-mail: henrik.koch@ntnu.no

## S1. QED Møller-Plesset perturbation theory

Consider an *ab initio* QED Hamiltonian $H$ able to model a molecular system interacting with a set $\{\alpha\}$ of non-interacting bosons.[1] This many-body Hamiltonian can be split in three terms

$$H = H_{mol} + \sum_{\alpha} H_{\alpha} + H_{int}, \tag{SE1}$$

where $H_{mol}$ is the molecular Hamiltonian, $H_{\alpha}$ is the $\alpha$-boson Hamiltonian and lastly $H_{int}$ is the interaction between the molecular system and the bosonic fields. Working in the Born-Oppenheimer approximation, the molecular Hamiltonian $H_{mol}$ reduces to the electronic Hamiltonian

$$H_{el} = \sum_{pq} h_{pq} E_{pq} + \frac{1}{2} \sum_{pqrs} g_{pqrs} e_{pqrs}, \tag{SE2}$$

where $h_{pq}$ and $g_{pqrs}$ are the one and two electron integrals. The second quantization formalism is adopted, so

$$E_{pq} = \sum_\sigma a^\dagger_{p\sigma} a_{q\sigma} \tag{SE3}$$

$$e_{pqrs} = E_{pq} E_{rs} - \delta_{rq} E_{ps},$$

with $a^\dagger_{p\sigma}$ and $a_{p\sigma}$ respectively creating and annihilating an electron in the orbital $p$ with spin $\sigma$. On the other hand, each field Hamiltonian $H_\alpha$ is written as a collection of quantum harmonic oscillators

$$H_\alpha = \sum_{k_\alpha} \omega_{k_\alpha} b^\dagger_{k_\alpha} b_{k_\alpha}, \tag{SE4}$$

where $k$ runs over boson modes and $b_{k,\alpha}$, $b^\dagger_{k,\alpha}$ do annihilate, create bosons of frequency $\omega_{k,\alpha}$. These harmonic oscillators are generally coupled linearly with the electronic degrees of freedom

$$H_{int} = \sum_{k_\alpha} \sum_{pq} g^{k_\alpha}_{pq} E_{pq} (b_{k_\alpha} + b^\dagger_{k_\alpha}), \tag{SE5}$$

where $g^{k_\alpha}_{pq}$ are coupling constants that can be in any order of accuracy in the multipolar expansion of the interaction.

This Hamiltonian

$$H = H_{el} + \sum_{k_\alpha} \omega_{k_\alpha} b^\dagger_{k_\alpha} b_{k_\alpha} + \sum_{k_\alpha} \sum_{pq} g^{k_\alpha}_{pq} E_{pq} (b_{k_\alpha} + b^\dagger_{k_\alpha}) \tag{SE6}$$

is defined in a space

$$\mathcal{H} = \mathcal{H}_{el} \bigotimes_\alpha \mathcal{H}_\alpha, \tag{SE7}$$

where $\mathcal{H}_{el}$ is the electronic Hilbert space and $\mathcal{H}_\alpha$ is the $\alpha$-boson Hilbert space. This tensor space is spanned by the states

$$\{|\mu\rangle \bigotimes_\alpha U^{coh}_\alpha |n_{1_\alpha}, ..., n_{k_\alpha}, ...\rangle\} \tag{SE8}$$

where $|\mu\rangle$ is an electronic occupation number state (ONS) and $|n_{1_\alpha}, ..., n_{k_\alpha}, ...\rangle$ is an $\alpha$-bosonic ONS. For the electronic ONSs $|\mu\rangle$, the quasiparticle formalism is adopted by considering as a reference the Hartree-Fock state

$$|\text{HF}\rangle = \prod_{i,\sigma}^{n_{occ}} a_{i\sigma}^\dagger |vac\rangle, \quad (\text{SE9})$$

where the low-lying molecular orbitals of the electronic vacuum $|vac\rangle$ are occupied. So, the $|\mu\rangle$ states are defined as excitations of the reference $|\text{HF}\rangle$

$$|\mu\rangle = \tau_\mu |\text{HF}\rangle, \quad (\text{SE10})$$

where $\tau_\mu$ is an excitation operator and considering $\tau_0 = I$, the identity. For the $\alpha$-boson ONSs, we consider as a reference the vacuum

$$|0_\alpha\rangle = |0_{1_\alpha}, ..., 0_{k_\alpha}, ...\rangle. \quad (\text{SE11})$$

So, the $|n_{1_\alpha}, ..., n_{k_\alpha}, ...\rangle$ are defined as excitations on this vacuum

$$|n_{1_\alpha}, ..., n_{k_\alpha}, ...\rangle = \prod_{k_\alpha} \frac{1}{\sqrt{n_{k_\alpha}!}} (b_{k_\alpha}^\dagger)^{n_{k_\alpha}} |0_\alpha\rangle, \quad (\text{SE12})$$

where the set $\{n_{\kappa_\alpha}\}$ are the occupation numbers of the $\alpha$-boson modes. The $U_\alpha^{coh}$ in eq. (SE8) are coherent-states transformations of the form

$$U_\alpha^{coh} = \prod_{k_\alpha} \exp\left(-z_{k_\alpha}(b_{k_\alpha} - b_{k_\alpha}^\dagger)\right), \quad (\text{SE13})$$

where $z_{k_\alpha}$ is a coherent-state parameters. We can change the quantum picture by passing in

the coherent-state basis for each boson and applying all the $U_\alpha^{coh}$ to the Hamiltonian

$$H_{coh} = \prod_\alpha U_\alpha^{coh\,\dagger}\, H\, U_\alpha^{coh}. \tag{SE14}$$

In this picture, the Hamiltonian is defined in a space spanned by the states

$$\{|\mu\rangle \bigotimes_\alpha |n_{1_\alpha}, ..., n_{k_\alpha}, ...\rangle \equiv |\mu, \{n_{k_\alpha}\}\rangle\}. \tag{SE15}$$

Due to the correlation between the particles of the many-body system, the exact eigenstates of $H_{coh}$ are complicated to model and most likely unknown in a finite order expansion of the basis in eq. (SE15). For this reason we can rely on Rayleigh-Shrödinger perturbation theory and split $H_{coh}$ in a zeroth-order solvable part $H_{coh}^{(0)}$ and a fluctuation potential $V$ where all the correlation effects do reside

$$H_{coh} = H_{coh}^{(0)} + \gamma V. \tag{SE16}$$

The $\gamma$ parameter is introduced in order to keep track of the perturbation orders of the expansion throughout the derivation. Following the Møller-Plesset (MP) scheme for electronic structure theory,[2] in the zeroth-order Hamiltonian we can consider the Fock operator coming from a mean-field treatment of the correlation effects

$$F_{coh} = \sum_{pq} F_{pq}^{coh} E_{pq}, \tag{SE17}$$

where $F_{pq}^{coh}$ are Fock matrix elements. Due to the presence of the bosons in the overall system, it is wise to add the harmonic oscillators in $H_{coh}^{(0)}$ as well

$$H_{coh}^{(0)} = \sum_{pq} F_{pq}^{coh} E_{pq} + \sum_{k_\alpha} \omega_{k_\alpha} b_{k_\alpha}^\dagger b_{k_\alpha}. \tag{SE18}$$

The states in eq. (SE15) are eigenfunctions of (SE18)

$$H_{coh}^{(0)} |\mu, \{n_{k_\alpha}\}\rangle^{(0)} = E_{\mu,\{n_{k_\alpha}\}}^{(0)} |\mu, \{n_{k_\alpha}\}\rangle^{(0)} \tag{SE19}$$

with eigenvalues

$$E_{\mu,\{n_{k_\alpha}\}}^{(0)} = E_{\text{HF},\{0_{k_\alpha}\}}^{(0)} + \epsilon_\mu + \sum_{k_\alpha} n_{k_\alpha} \omega_{k_\alpha}, \tag{SE20}$$

where $E_{\text{HF},\{0_{k_\alpha}\}}$ is the ground state zeroth order energy and $\epsilon_\mu$ is the excitation energy between the the electronic $|\text{HF}\rangle$ and $|\mu\rangle$ Slater determinants. To obtain eq. (SE20) we made use of the diagonal form of the Fock matrix in the canonical basis. The eigenstates $|\mu, \{n_{k_\alpha}\}\rangle^{(0)}$ are the ones spanning the coupled electron-bosons Hilbert space where the coherent-state transformed Hamilotnian is defined (eq. (SE15)). The fluctuation potential $V$ is straightforwardly defined as

$$V = H_{coh} - H_{coh}^{(0)}. \tag{SE21}$$

In order to see how the states in eq. (SE15) are affected by the presence of the perturbation $V$, we expand them perturbatively in $\gamma$

$$|\psi_{\mu,\{n_{k_\alpha}\}}\rangle = |\mu, \{n_{k_\alpha}\}\rangle^{(0)} + \gamma |\mu, \{n_{k_\alpha}\}\rangle^{(1)} + \gamma^2 |\mu, \{n_{k_\alpha}\}\rangle^{(2)} + ... \tag{SE22}$$

and the same we do for the associated energies

$$E_{\mu,\{n_{k_\alpha}\}} = E_{\mu,\{n_{k_\alpha}\}}^{(0)} + \gamma E_{\mu,\{n_{k_\alpha}\}}^{(1)} + \gamma^2 E_{\mu,\{n_{k_\alpha}\}}^{(2)} + ... \tag{SE23}$$

The terms in zeroth-order with respect to $\gamma$ are the ones showing in eq. (SE19) and (SE20). Now, we can insert eqs. (SE22) and (SE23) in the Shrödinger equation for the full Hamiltonian

$$H_{coh} |\psi_{\mu,\{n_{k_\alpha}\}}\rangle = E_{\mu,\{n_{k_\alpha}\}} |\psi_{\mu,\{n_{k_\alpha}\}}\rangle \tag{SE24}$$

and obtain

$$(H_{coh}^{(0)} + \gamma V)(|\mu, \{n_{k_\alpha}\}\rangle^{(0)} + \gamma |\mu, \{n_{k_\alpha}\}\rangle^{(1)} + \gamma^2 |\mu, \{n_{k_\alpha}\}\rangle^{(2)} + ...) =$$
$$= (E_{\mu,\{n_{k_\alpha}\}}^{(0)} + \gamma E_{\mu,\{n_{k_\alpha}\}}^{(1)} + \gamma^2 E_{\mu,\{n_{k_\alpha}\}}^{(2)} + ...)(|\mu, \{n_{k_\alpha}\}\rangle^{(0)} + \gamma |\mu, \{n_{k_\alpha}\}\rangle^{(1)} + \gamma^2 |\mu, \{n_{k_\alpha}\}\rangle^{(2)} + ...). \tag{SE25}$$

We notice that the equation must hold in each order $\gamma^n$ of the perturbation expansion:

$$\gamma^0: \quad H_{coh}^{(0)} |\mu, \{n_{k_\alpha}\}\rangle^{(0)} = E_{\mu,\{n_{k_\alpha}\}}^{(0)} |\mu, \{n_{k_\alpha}\}\rangle^{(0)} \tag{SE26}$$

$$\gamma^1: \quad H_{coh}^{(0)} |\mu, \{n_{k_\alpha}\}\rangle^{(1)} + V |\mu, \{n_{k_\alpha}\}\rangle^{(0)} =$$
$$= E_{\mu,\{n_{k_\alpha}\}}^{(0)} |\mu, \{n_{k_\alpha}\}\rangle^{(1)} + E_{\mu,\{n_{k_\alpha}\}}^{(1)} |\mu, \{n_{k_\alpha}\}\rangle^{(0)} \tag{SE27}$$

$$\gamma^2: \quad H_{coh}^{(0)} |\mu, \{n_{k_\alpha}\}\rangle^{(2)} + V |\mu, \{n_{k_\alpha}\}\rangle^{(1)} =$$
$$= E_{\mu,\{n_{k_\alpha}\}}^{(0)} |\mu, \{n_{k_\alpha}\}\rangle^{(2)} + E_{\mu,\{n_{k_\alpha}\}}^{(1)} |\mu, \{n_{k_\alpha}\}\rangle^{(1)} + E_{\mu,\{n_{k_\alpha}\}}^{(2)} |\mu, \{n_{k_\alpha}\}\rangle^{(0)} \tag{SE28}$$

...

At the zeroth-order in (SE26) we have the Shrödinger equation in (SE19).

For the first-order in (SE27) we can expand $|\mu, \{n_{k_\alpha}\}\rangle^{(1)}$ in the zeroth-order basis

$$|\mu, \{n_{k_\alpha}\}\rangle^{(1)} = \sum_{\nu, \{n_{k_\beta}\}} c_{\mu,\{n_{k_\alpha}\};\ \nu,\{n_{k_\beta}\}}^{(1)} |\nu, \{n_{k_\beta}\}\rangle^{(0)} \tag{SE29}$$

and project on $^{(0)}\langle\mu,\{n_{k_\alpha}\}|$ to obtain the first-order energy correction

$$E^{(1)}_{\mu,\{n_{k_\alpha}\}} = {}^{(0)}\langle\mu,\{n_{k_\alpha}\}|V|\mu,\{n_{k_\alpha}\}\rangle^{(0)}. \tag{SE30}$$

On the other hand, if we project on a generic $^{(0)}\langle\rho,\{n_{k_\theta}\}|$ different from $^{(0)}\langle\mu,\{n_{k_\alpha}\}|$, we obtain the expression for the expansion coefficients in eq. (SE29)

$$c^{(1)}_{\mu,\{n_{k_\alpha}\};\,\nu,\{n_{k_\beta}\}} = -\frac{{}^{(0)}\langle\nu,\{n_{k_\beta}\}|V|\mu,\{n_{k_\alpha}\}\rangle^{(0)}}{E^{(0)}_{\nu,\{n_{k_\beta}\}} - E^{(0)}_{\mu,\{n_{k_\alpha}\}}} \quad ; \quad \nu,\{n_{k_\beta}\} \neq \mu,\{n_{k_\alpha}\}. \tag{SE31}$$

To obtain eq. (SE30) and eq. (SE31) we made use of the orthogonality of the zeroth-order basis and of eq. (SE26). So, the first-order correction of the basis eigenstates is

$$|\mu,\{n_{k_\alpha}\}\rangle^{(1)} = |\mu,\{n_{k_\alpha}\}\rangle^{(0)} - \sum_{\nu,\{n_{k_\beta}\}} \frac{{}^{(0)}\langle\nu,\{n_{k_\beta}\}|V|\mu,\{n_{k_\alpha}\}\rangle^{(0)}}{E^{(0)}_{\nu,\{n_{k_\beta}\}} - E^{(0)}_{\mu,\{n_{k_\alpha}\}}} |\nu,\{n_{k_\beta}\}\rangle^{(0)}. \tag{SE32}$$

For the second-order in (SE28) we can again expand $|\mu,\{n_{k_\alpha}\}\rangle^{(2)}$ in the zeroth-order basis

$$|\mu,\{n_{k_\alpha}\}\rangle^{(2)} = \sum_{\nu,\{n_{k_\beta}\}} c^{(2)}_{\mu,\{n_{k_\alpha}\};\,\nu,\{n_{k_\beta}\}} |\nu,\{n_{k_\beta}\}\rangle^{(0)} \tag{SE33}$$

and project on $^{(0)}\langle\mu,\{n_{k_\alpha}\}|$ to obtain the second-order energy correction

$$E^{(2)}_{\mu,\{n_{k_\alpha}\}} = -\sum_{\nu,\{n_{k_\beta}\}} \frac{{}^{(0)}\langle\nu,\{n_{k_\beta}\}|V|\mu,\{n_{k_\alpha}\}\rangle^{(0)}\,{}^{(0)}\langle\mu,\{n_{k_\alpha}\}|V|\nu,\{n_{k_\beta}\}\rangle^{(0)}}{E^{(0)}_{\nu,\{n_{k_\beta}\}} - E^{(0)}_{\mu,\{n_{k_\alpha}\}}} \quad ; \quad \nu,\{n_{k_\beta}\} \neq \mu,\{n_{k_\alpha}\}. \tag{SE34}$$

Yet again, with the same technique used before we can determine the second-order correction to the eigenstates.

Iterating the procedure we can obtain the energy and eigenstates corrections in all orders of the Møller-Plesset perturbation hierarchy with the well known $2n + 1$ rule.

## S2. Strong coupling QED-MP2

We consider the single-mode Pauli-Fierz Hamiltonian modeling the light-matter interaction between a molecular system and a cavity photon[3]

$$H = H_e + \omega b^\dagger b + \frac{\lambda^2}{2}(\mathbf{d} \cdot \boldsymbol{\epsilon})^2 - \lambda\sqrt{\frac{\omega}{2}}(\mathbf{d} \cdot \boldsymbol{\epsilon})(b^\dagger + b) \tag{SE35}$$

where $b$ and $b^\dagger$ annihilate and create a photon of frequency $\omega$ and with polarization $\boldsymbol{\epsilon}$. This Hamiltonian is in the length gauge, with

$$\mathbf{d} = \sum_{pq} \mathbf{d}_{pq} E_{pq} \tag{SE36}$$

being the molecular dipole operator. The $\lambda$ parameter represents the light-matter coupling strength. For the ground-state wave function parametrization we consider the SC-QED-HF Ansatz[4,5]

$$|\psi\rangle = \exp\left(-\frac{\lambda}{\sqrt{2\omega}} \sum_p \eta_p \tilde{E}_{pp}(b - b^\dagger)\right) |\text{HF}, 0\rangle, \tag{SE37}$$

where the electrons are dressed with cavity photons by means of the orbital specific coherent-state transformation

$$U_{\text{SC}} = \exp\left(-\frac{\lambda}{\sqrt{2\omega}} \sum_p \eta_p \tilde{E}_{pp}(b - b^\dagger)\right). \tag{SE38}$$

The tilde $\sim$ symbol denotes integrals and operators in the basis that diagonalize $(\mathbf{d} \cdot \boldsymbol{\epsilon})$:

$$\sum_{rs} C_{rp}(\mathbf{d} \cdot \boldsymbol{\epsilon})_{rs} C_{sq} = (\tilde{\mathbf{d}} \cdot \boldsymbol{\epsilon})_{pp} \delta_{pq}, \tag{SE39}$$

where $\mathbf{C}$ is an orthonormal rotation matrix connecting molecular and dipole orbitals

$$\mathbf{d} = \sum_p \tilde{\mathbf{d}}_{pp} \tilde{E}_{pp}. \tag{SE40}$$

Now we apply the QED Møller-Plesset perturbation theory developed in the previous section to the Hamiltonian (SE35) and the wave function (SE37). Proceeding, we first change the quantum picture by SC-transforming the Pauli-Fierz Hamiltonian

$$H_{\text{SC}} = U_{\text{SC}}^\dagger \, H \, U_{\text{SC}}$$
$$= \sum_{pq} \tilde{h}_{pq}^{\text{SC}} Y_{pq} \tilde{E}_{pq} + \frac{1}{2} \sum_{pqrs} \tilde{g}_{pqrs}^{\text{SC}} Y_{pqrs} \tilde{e}_{pqrs} \qquad \text{(SE41)}$$
$$+ \omega b^\dagger b - \lambda \sqrt{\frac{\omega}{2}} \sum_p ((\tilde{\mathbf{d}} \cdot \boldsymbol{\epsilon})_{pp} - \eta_p) \tilde{E}_{pp}(b + b^\dagger),$$

where the redefined SC one and two electron integrals are

$$\tilde{h}_{pq}^{\text{SC}} = \tilde{h}_{pq} + \frac{\lambda^2}{2}((\tilde{\mathbf{d}} \cdot \boldsymbol{\epsilon})_{pp} - \eta_p)^2 \delta_{pq} \qquad \text{(SE42)}$$

$$\tilde{g}_{pqrs}^{\text{SC}} = \tilde{g}_{pqrs} + \lambda^2 ((\tilde{\mathbf{d}} \cdot \boldsymbol{\epsilon})_{pp} - \eta_p)((\tilde{\mathbf{d}} \cdot \boldsymbol{\epsilon})_{rr} - \eta_r) \delta_{pq} \delta_{rs}, \qquad \text{(SE43)}$$

while the $Y_{pq}$ and $Y_{pqrs}$ photonic operators are defined as follows

$$Y_{pq} = \exp\left(\frac{\lambda}{\sqrt{2\omega}}(\eta_p - \eta_q)(b - b^\dagger)\right) \qquad \text{(SE44)}$$

$$Y_{pqrs} = \exp\left(\frac{\lambda}{\sqrt{2\omega}}(\eta_p - \eta_q + \eta_r - \eta_s)(b - b^\dagger)\right). \qquad \text{(SE45)}$$

The Hamiltonian in eq. (SE41) is defined in a space spanned by the states

$$\{|\mu, n\rangle\} \qquad \text{(SE46)}$$

with $|\text{HF}, 0\rangle$ being the optimized ground state. We define as the SC zeroth-order Hamiltonian

$$H_{\text{SC}}^{(0)} = \sum_{pq} F_{pq}^{\text{SC}} E_{pq} + \omega b^\dagger b, \qquad \text{(SE47)}$$

where the SC-Fock matrix elements in the dipole basis read

$$\tilde{F}^{\rm SC}_{pq} = \tilde{h}^{\rm SC}_{pq} Q_{pq} + \frac{1}{2} \sum_{rs} (2\tilde{g}^{\rm SC}_{pqrs} - \tilde{g}^{\rm SC}_{psrq}) Q_{pqrs} \tilde{D}_{rs}, \tag{SE48}$$

$\tilde{D}_{pq}$ are elements of the one electron density matrix (in the dipole basis as well)

$$\tilde{D}_{pq} = \langle {\rm HF}| \tilde{E}_{pq} |{\rm HF}\rangle \tag{SE49}$$

and the Gaussian factors carrying the $\omega$-correlation are

$$Q_{pq} = \langle Y_{pq}\rangle_0 = \exp\left(-\frac{\lambda^2}{4\omega}(\eta_p - \eta_q)^2\right) \tag{SE50}$$

$$Q_{pqrs} = \langle Y_{pqrs}\rangle_0 = \exp\left(-\frac{\lambda^2}{4\omega}(\eta_p - \eta_q + \eta_r - \eta_s)^2\right). \tag{SE51}$$

The Fock matrix elements in eqs. (SE47) and (SE48) are connected by

$$\sum_{rs} C_{rp} F_{rs} C_{sq} = \tilde{F}_{pq}, \tag{SE52}$$

where the canonical to dipole basis transformation $\mathbf{C}$ is defined in eq. (SE39).

The states in eq. (SE46) are eigenfunctions of (SE47)

$$H^{(0)}_{\rm SC} |\mu, n\rangle^{(0)} = E^{(0)}_{\mu n} |\mu, n\rangle^{(0)} \tag{SE53}$$

with eigenvalues

$$E^{(0)}_{\mu n} = E^{(0)}_{{\rm HF},0} + \epsilon_\mu + n\omega. \tag{SE54}$$

As discussed in the previous section, the states $|\mu, n\rangle^{(0)}$ are the ones spanning the light-matter Hilbert space where the SC-trasformed Pauli-Fierz Hamiltonian is defined (eq. (SE46)). For this reason, from now on, we drop the zeroth order index for the eigenstates. For $\mu = 0$ the electronic excitation is trivially $\epsilon_0 = 0$ a.u. On the other hand, for excited determinants, we

have for singles ($\mu \in S$), doubles ($\mu \in D$) and so on

$$\epsilon_\mu = \epsilon_a - \epsilon_i, \quad \text{for} \quad |\mu\rangle = E_{ai}|\text{HF}\rangle \tag{SE55}$$

$$\epsilon_\mu = \epsilon_a - \epsilon_i + \epsilon_b - \epsilon_j, \quad \text{for} \quad |\mu\rangle = E_{bj}E_{ai}|\text{HF}\rangle \tag{SE56}$$

...

where $i, j, ...$ and $a, b, ...$ label respectively occupied and virtual orbitals in the $|\text{HF}\rangle$ state. Then, the perturbation $V = H_{\text{SC}} - H_{\text{SC}}^{(0)}$ reads

$$\begin{aligned}
V = &\sum_{pq} \tilde{h}_{pq}^{\text{SC}}(Y_{pq} - Q_{pq})\tilde{E}_{pq} + \frac{1}{2}\sum_{pqrs} \tilde{g}_{pqrs}^{\text{SC}} Y_{pqrs} \tilde{e}_{pqrs} \\
&- \frac{1}{2}\sum_{pqrs}(2\tilde{g}_{pqrs}^{\text{SC}} - \tilde{g}_{psrq}^{\text{SC}})Q_{pqrs}\tilde{D}_{rs}\tilde{E}_{pq} \\
&- \lambda\sqrt{\frac{\omega}{2}}\sum_p((\tilde{\mathbf{d}}\cdot\boldsymbol{\epsilon})_{pp} - \eta_p)\tilde{E}_{pp}(b + b^\dagger).
\end{aligned} \tag{SE57}$$

The sum of the zeroth and first-order energy for every state in (SE46) corresponds to the energy associated to that state calculated at the mean-field level of theory. The first correction to these energies happen to be in the second order of the QED Møller-Plesset perturbation theory. With the focus on building up correlation on top of the SC-QED-HF ground-state, the QED-MP2 correction reads

$$E_{\text{HF},0}^{(2)} = -\sum_{\mu,n} \frac{\langle \mu, n | H_{\text{SC}} | \text{HF}, 0\rangle \langle \text{HF}, 0 | H_{\text{SC}} | \mu, n\rangle}{E_{\mu,n}^{(0)} - E_{\text{HF},0}^{(0)}} \quad ; \quad \mu, n \neq \text{HF}, 0. \tag{SE58}$$

In this last equation we substituted $V$ with $H_{\text{SC}}$ because of eq. (SE53) and the orthogonality of the basis states in eq. (SE46). Because of the hermicity of the Hamiltonian, we can rewrite this correction in the following manner

$$E_{\text{HF},0}^{(2)} = -\sum_{\mu,n} \frac{|\langle \text{HF}, 0 | H_{\text{SC}} | \mu, n\rangle|^2}{E_{\mu,n}^{(0)} - E_{\text{HF},0}^{(0)}} \quad ; \quad \mu, n \neq \text{HF}, 0. \tag{SE59}$$

Now we go through each type of excitation that contributes to the correction in eq. (SE59) and, for each of them, we see which terms of the SC-Hamiltonian in eq. (SE41) do contribute. The photon energy $\omega b^\dagger b$ obviously never contributes, while for evaluation of the denominators we use eq. (SE54).

1. We first start by considering purely photonic excitations ($\mu = 0$, $n > 0$). The $n = 1$ contribution is zero because of the Brillouin condition in the photonic part

$$\langle \mathrm{HF}, 0 | \left[ H_{\mathrm{SC}}, \tilde{E}_{pp}(b - b^\dagger) \right] | \mathrm{HF}, 0 \rangle = 0. \tag{SE60}$$

The SC-transformed electronic-like Hamiltonian

$$H_{el}^{\mathrm{SC}} = \sum_{pq} \tilde{h}_{pq}^{\mathrm{SC}} Y_{pq} \tilde{E}_{pq} + \frac{1}{2} \sum_{pqrs} \tilde{g}_{pqrs}^{\mathrm{SC}} Y_{pqrs} \tilde{e}_{pqrs} \tag{SE61}$$

contributes for all the $n > 1$. Specifically

$$\sum_{n=2}^{\infty} \frac{|\langle \mathrm{HF}, 0 | H_{el}^{\mathrm{SC}} | \mathrm{HF}, n \rangle|^2}{E_{\mathrm{HF},n}^{(0)} - E_{\mathrm{HF},0}^{(0)}} = \sum_{n=2}^{\infty} \frac{(E^n)^2}{n\omega}, \tag{SE62}$$

where

$$E^n = 2 \sum_i h_{ii}^n + \sum_{ij} (2 g_{iijj}^n - g_{ijji}^n) \tag{SE63}$$

and the the nth one and two electron integrals in the canonical basis depend from the Laguerre polynomials obtained by calculation of the displacement Franck-Condon factors

$$h_{pq}^n = \frac{1}{\sqrt{n!}} \sum_{rs} C_{pr} \, \tilde{h}_{rs}^{\mathrm{SC}} \, Q_{rs} \left( \frac{\lambda}{\sqrt{2\omega}} (\eta_r - \eta_s) \right)^n C_{qs} \tag{SE64}$$

$$g_{pqrs}^n = \frac{1}{\sqrt{n!}} \sum_{tuvz} C_{pt} C_{rv} \, \tilde{g}_{tuvz}^{\mathrm{SC}} \, Q_{tuvz} \left( \frac{\lambda}{\sqrt{2\omega}} (\eta_t - \eta_u + \eta_v - \eta_z) \right)^n C_{qu} C_{sz}. \tag{SE65}$$

2. Secondly, we consider single (S) excitations in the orbital space coupled with generic photonic excitations ($\mu \in S$, $n > 0$). The $n = 0$ contributions are zero because of the

S-12

Brillouin condition in the orbital part

$$\langle \mathrm{HF}, 0 | \left[ H_{\mathrm{SC}}, E_{ai}^{-} \right] | \mathrm{HF}, 0 \rangle = 0. \tag{SE66}$$

Using Slater-Condon rules for single excited determinants, we obtain

$$\sum_{\mu \in S} \sum_{n=1}^{\infty} \frac{|\langle \mathrm{HF}, 0 | H_{\mathrm{SC}} | \mu, n \rangle|^2}{E_{\mu,n}^{(0)} - E_{\mathrm{HF},0}^{(0)}} = \sum_{ai} \sum_{n=1}^{\infty} \frac{(F_{ai}^n)^2}{n\omega + \epsilon_a - \epsilon_i}, \tag{SE67}$$

where the nth-Fock matrix elements read

$$F_{pq}^n = h_{pq}^n + \sum_{i} (2g_{pqii}^n - g_{piiq}^n) + \delta_{n1} \lambda \sqrt{\frac{\omega}{2}} ((\tilde{\mathbf{d}} \cdot \boldsymbol{\epsilon})_{pp} - \eta_p). \tag{SE68}$$

Again we made use of the displacement Franck-Condon factors:

$$\langle n | e^{-\alpha(b-b^\dagger)} | m \rangle = \begin{cases} \sqrt{\frac{m!}{n!}} \, \alpha^{n-m} e^{-\alpha^2/2} L_m^{n-m}(\alpha^2), & n \geq m \\[2ex] \sqrt{\frac{n!}{m!}} \, (-\alpha)^{m-n} e^{-\alpha^2/2} L_n^{m-n}(\alpha^2), & n < m \end{cases} \tag{SE69}$$

where $L_p^q$ is the Laguerre $q$-th order polynomial of degree $p$.

3. Lastly, we consider doubly (D) excitations in the orbital space coupled with generic photonic excitations ($\mu \in D$, $n > 0$). In this case, all the $n > 0$ do contribute. No more than double excitation need to be considered because, according to Slater-Condon rules, the SC-Hamiltonian in eq. (SE41) cannot connect states differing by more than two occupied orbitals. Only the two electron part of the SC-transformed electron-like Hamiltonian in eq. (SE61) do contribute. Analogously to standard MP2 theory, this correction reads

$$\sum_{\mu \in D} \sum_{n=0}^{\infty} \frac{|\langle \mathrm{HF}, 0 | H_{\mathrm{SC}} | \mu, n \rangle|^2}{E_{\mu,n}^{(0)} - E_{\mathrm{HF},0}^{(0)}} = \sum_{aibj} \sum_{n=0}^{\infty} \frac{g_{aibj}^n (2g_{aibj}^n - g_{ajbi}^n)}{n\omega + \epsilon_a - \epsilon_i + \epsilon_b - \epsilon_j}. \tag{SE70}$$

So, the second order energy correction to the SC ground-state is

$$E^{(2)} = -\sum_{n=2}^{\infty} \frac{(E^n)^2}{n\omega} - \sum_{n=1}^{\infty}\sum_{ai} \frac{(F_{ai}^n)^2}{n\omega + \epsilon_a - \epsilon_i} - \sum_{n=0}^{\infty}\sum_{aibj} \frac{g_{aibj}^n(2g_{aibj}^n - g_{ajbi}^n)}{n\omega + \epsilon_a + \epsilon_b - \epsilon_i - \epsilon_j}. \quad \text{(SE71)}$$

By capturing electron-photon correlation, this MP2 correction is correctly non size-extensive when considering two subsystems largely separated within a cavity.

# References

(1) Berestetskii, V. B.; Lifshitz, E. M.; Pitaevskii, L. P. *Quantum Electrodynamics: Volume 4*; Butterworth-Heinemann, 1982; Vol. 4.

(2) Helgaker, T.; Jorgensen, P.; Olsen, J. *Molecular electronic-structure theory*; John Wiley & Sons, 2013.

(3) Cohen-Tannoudji, C.; Dupont-Roc, J.; Grynberg, G. *Photons and atoms: introduction to quantum electrodynamics*; John Wiley & Sons, 2024.

(4) Riso, R. R.; Haugland, T. S.; Ronca, E.; Koch, H. Molecular orbital theory in cavity QED environments. *Nat. Commun.* **2022**, *13*, 1368.

(5) El Moutaoukal, Y.; Riso, R. R.; Castagnola, M.; Koch, H. Toward polaritonic molecular orbitals for large molecular systems. *J. Chem. Theory Comput.* **2024**, *20*, 8911–8920.



\end{document}